\documentclass[%
 reprint,
 amsmath,amssymb,
 aps,
]{revtex4-2}
\usepackage[english]{babel} 
\usepackage{graphicx}% Include figure files
\usepackage{dcolumn}% Align table columns on decimal point
\usepackage{bm}% bold math
\usepackage{hyperref}% add hypertext capabilities
\usepackage{svg}
\usepackage{graphicx}
\usepackage{caption}
\usepackage{subcaption}
\usepackage{siunitx}
\usepackage{pgf}
\begin{document}

%\preprint{APS/123-QED}

\title{Engineering optical forces through Maxwell stress tensor inverse design}% Force line breaks with \\
\author{Beñat Martinez de Aguirre Jokisch}
 \email{bmdaj@dtu.dk}
\author{Rasmus Ellebæk Christiansen}%
\author{Ole Sigmund}%

\affiliation{%
 Department of Civil and Mechanical Engineering,Technical University of Denmark, Nils Koppels Allé Building 404, Kongens Lyngby, 2800, Denmark%\\
}%
\affiliation
{NanoPhoton - Center for Nanophotonics, Technical University of Denmark, Ørsteds Plads 345A, Kongens Lyngby, 2800, Denmark}

\date{\today}% It is always \today, today,

\begin{abstract}
Precise spatial manipulation of particles via optical forces is essential in many research areas, ranging from biophysics to atomic physics. Central to this effort is the challenge of designing optical systems that are optimized for specific applications. Traditional design methods often rely on trial-and-error methods, or on models that approximate the particle as a point dipole, which only works for particles much smaller than the wavelength of the electromagnetic field. In this work, we present a general inverse design framework based on the Maxwell stress tensor formalism capable of simultaneously designing all components of the system, while being applicable to particles of arbitrary sizes and shapes.  Notably, we show that with small modifications to the baseline formulation, it is possible to engineer systems capable of attracting, repelling, accelerating, oscillating, and trapping particles. We demonstrate our method using various case studies where we simultaneously design the particle and its environment, with particular focus on free-space particles and particle-metalens systems.
%\begin{description}
%Secondary publications and %information retrieval purposes.
%You may use the \texttt{description} %environment to structure your abstract;
% the optional argument of the \verb+\item+ command to give the category of each item. 
%\end{description}
\end{abstract}
\maketitle
%\tableofcontents
\onecolumngrid
\section{ Introduction}
Optical forces are a result of an exchange of momentum in the interaction of electromagnetic fields with absorbing or scattering matter. These forces, which are usually in the order of piconewtons to femtonewtons are inconsequential in macroscopic applications, but become non-negligible for the motion of micro- and nano-scale particles, scales at which  the optical forces can compete with gravitational forces. By ingenious  use of these optical forces it becomes possible to engineer the motion of nanoscopic and microscopic particles, leading to the experimental realization of optical trapping \cite{ashkin_acceleration_1970, moffitt_recent_2008}, optical cooling \cite{ashkin_optical_1997},  optical binding \cite{burns_optical_1989,dholakia_colloquium_2010}, and sorting  and transporting particles \cite{wang_microfluidic_2005,almaas_possible_2013}, among 
others. These techniques have been successfully applied in different research areas; such as atomic physics, to trap isolated atoms \cite{chang_trapping_2009}; biophysics, to realize single-molecule 
biophysics experiments \cite{bustamante_optical_2021}; or chemistry, to achieve single-molecule fluorescence spectroscopy \cite{li_evidence_2006}.\\

Most traditional setups designed to exert optical forces are based on free-space optical elements \cite{ashkin_acceleration_1970, ashkin_optical_1997}, which are macroscopic devices operating in the ray-optics regime \cite{ashkin_forces_1992}. Having the ability to miniaturize these systems using optical
nanostructures offers new possibilities for integration and energy efficiency; however, it is not straightforward how to design optimal nanostructures for different applications. Traditionally, the optimization of most of these setups relies on intuition and trial-and-error based 
approaches \cite{xiao_-chip_2023, brooks_shape-directed_2019}, but recently some works have started integrating numerical optimization techniques in their optical design schemes  \cite{Lee:17, nelson_inverse_2024, martinez_de_aguirre_jokisch_omnidirectional_2024}. In this work, we derive an inverse design formulation based on topology optimization (TopOpt), to design nanostructures and particles for different applications. TopOpt is a design optimization method widely used in the design of optical applications \cite{jensen_topology_2011} like cavities 
\cite{wang_maximizing_2018, yao_trace_2022,albrechtsen_nanometer-scale_2022}, 
microresonators \cite{ahn_photonic_2022}, metalenses \cite{chung_high-na_2020,christiansen_compact_2021, li_inverse_2022} and more. 
TopOpt has recently also been applied to optical trapping problems, with the design of plasmonic nanotweezers \cite{nelson_inverse_2024}
and integrated dielectric nanocavities \cite{martinez_de_aguirre_jokisch_omnidirectional_2024}. These optical force-based optimization formalisms modelled the optical forces acting on the trapped particle in the dipole approximation \cite{novotny_principles_2012}, where the particle size is much smaller than the wavelength of the electromagnetic field, and thus with negligible self-induced back-action (SIBA) or scattering forces (e.g., radiation pressure or spin-curl forces) \cite{bustamante_optical_2021, novotny_principles_2012}. Furthermore, these works only considered the optimization of the trapping device without optimizing the particle geometry, a key challenge in optomechanic design \cite{gluck, Rodriguez_2014}. In contrast, in this work, we derive an inverse design formulation based on the Maxwell stress tensor (MST) 
that is used to calculate the forces on particles of arbitrary sizes and shapes. To this end, we derive the 
adjoint sensitivity analysis to calculate the gradients of the figure of merit (FOM) with respect to the design variables, to enable the use of efficient gradient-based optimizers in our TopOpt scheme.\\

Subsequently, we apply the derived formalism to selected examples, where we optimize custom figures of merit for free-space 
 particle systems and metalens-particle systems. We choose metasurface and metalens systems for our demonstrations due to the modelling simplicity \cite{christiansen_compact_2021}, in combination with their great potential for many applications, such as optical trapping \cite{huang_metasurface_2023, xiao_-chip_2023}, and particle accelerators \cite{bar-lev_design_2019}. Inspired by many of these applications we simultaneously optimize the  design of the metalens and the particle for different objectives; such as, attracting and repelling the particle, or trapping the particle in a target point in space. The derivation of the proposed framework paves the way for further optimization of more complex optomechanical systems, such as the design of optically actuated mechanical devices \cite{ivanyi_optically_2024}, the design of novel active particles beyond self-propelling particles \cite{bechinger_active_2016}, novel optically-driven particle paths \cite{zemanek_perspective_2019, macdonald_microfluidic_2003, shilkin_directional_2017}, or many-body systems \cite{merrill_many-body_2009, chang_colloquium_2018, bechinger_active_2016}. Note that the base code developed for this framework is open-source and is readily available at: \url{www.topopt.dtu.dk } and  \url{https://github.com/bmdaj/MST_TopOpt}.

 \section{Analytical physics model and discretized optimization problem}

\begin{figure}[ht]
    \centering
    \begin{minipage}{0.9\linewidth}
        \hspace*{-1cm}
        \centering
        \includegraphics[scale=0.22]{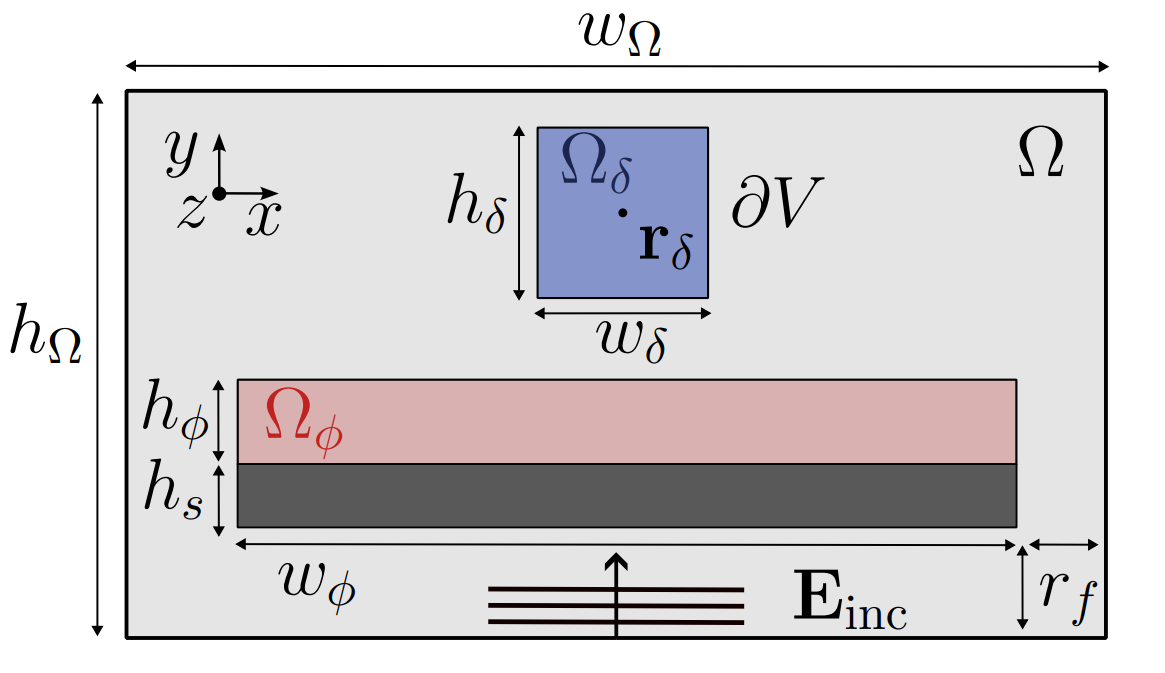}
        \caption{Two-dimensional model domain $\Omega$ with height $h_\Omega$ and
    width $w_\Omega$. In the model domain, there are two designable regions:
    the red region $\Omega_\phi$ is the metalens design region,
    which has a height $h_{\phi}$ and width $w_{\phi}$, and which is located on top of a substrate with height $h_s$; while the blue region
    $\Omega_\delta$ bounded by $\partial V$ is the particle design region centered around the point $\mathbf{r}_\delta$ with height  $h_{\delta}$
    and width $w_{\delta}$. The substrate is separated by a distance equal to the filter radius $r_f$ from the horizontal and vertical ends of the domain. A plane wave with an electric field $\mathbf{E}_\text{inc}$ excites the model domain from the lower edge of the simulation domain.}
        \label{fig:1}
    \end{minipage}
\end{figure}

In our examples we model the electromagnetic fields and the resulting optical forces, using the two-dimensional domain
$\Omega$ depicted in \autoref{fig:1}, that describes the metalens-particle system. However, our derivations are general and hold for three-dimensional systems. To calculate the electromagnetic fields we solve Maxwell's equations, where we
assume time-harmonic behavior. For simplicity, we also assume out-of-plane ($z$)
spatial invariance and	linear, static, homogeneous, isotropic, non-dispersive
and non-magnetic materials. Assuming out-of-plane polarization of the electric
field (TE polarization), Maxwell's equations can be combined into a
Helmholtz-type partial differential equation \cite{christiansen_compact_2021}:
\begin{equation}\label{eq:helmholtz}
    \nabla \cdot\left(\nabla E_z(\mathbf{r})\right)+k^2 \varepsilon_r(\mathbf{r})
    E_z(\mathbf{r})= f(\mathbf{r}), \quad \mathbf{r} \in \Omega \subset \mathbb{R}^2,
\end{equation}
where $E_z$ is the $z$-component of the electric field, $k$ is the wavenumber, $\varepsilon_r$ is the complex relative dielectric permittivity, and $f$ denotes the
forcing term, which is modeled as incident plane wave normal to the bottom
boundary with a field $\mathbf{E}_\text{inc} = E_0\, \text{e}^{-\mathrm{i}(k\cdot y)} \, \mathbf{n}_z$, where $E_0$ is the electric field amplitude, $\mathrm{i}$ is the imaginary unit, and $\mathbf{n}_z$ is a unitary vector in the $z$ direction. We apply first-order absorbing boundary conditions on all
the exterior boundaries:
\begin{equation}\label{eq:first_abs}
    \mathbf{n} \cdot \nabla E_z(\mathbf{r})=-\mathrm{i} k E_z(\mathbf{r}), \quad
    \mathbf{r} \in \Gamma \subset \Omega
\end{equation}
where $\mathbf{n}$ denotes the unitary vector normal to the boundary $\Gamma$. For TE polarization the rest of the
electric field components, and the out-of-plane component of the magnetic
field are zero ($E_x=E_y=H_z=0$), while the in-plane magnetic fields
($H_x$, $H_y$) can be calculated from Maxwell's equations:
\begin{equation} \label{eq:mag_fields}
    H_x = \frac{\mathrm{i}}{\omega \mu_0} \frac{\partial E_z}{\partial y}
    ,\quad \quad
    H_y = -\frac{\mathrm{i}}{\omega \mu_0} \frac{\partial E_z}{\partial x}\,.
\end{equation}
where $\omega$ is the angular frequency, and $\mu_0$ is the free-space
permeability. The model is discretized and solved using the finite-element
method using bi-linear quadratic elements \cite{jin_finite_2002}.\\

\begin{figure}[ht]
    
        %\hspace*{-1.25cm}
        %\vspace*{0.15cm}
        \centering
        \begin{minipage}{0.9\linewidth}
        \hspace*{-1cm}

        \centering
        \includegraphics[scale=0.2]{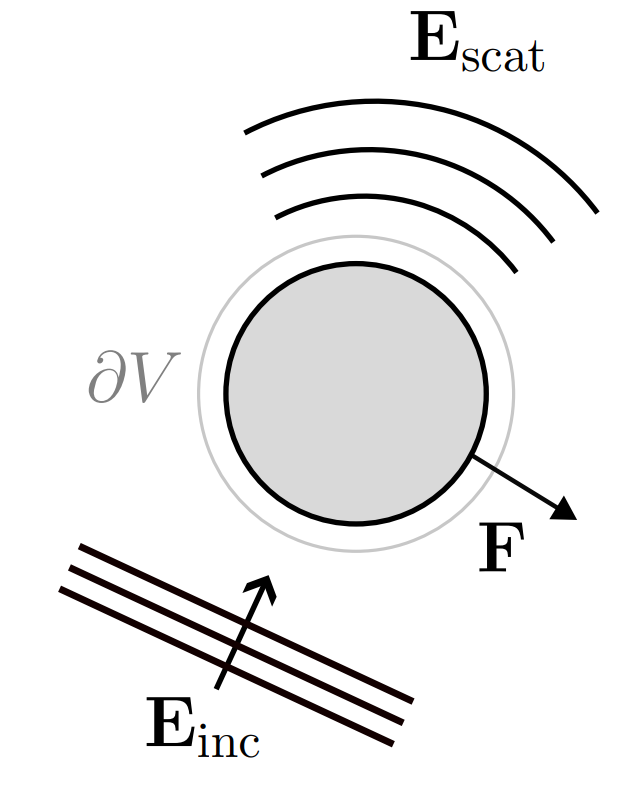}
        \caption{A scattering
    particle enclosed by a boundary $\partial V$ is excited by an incident field $\mathbf{E}_\text{inc}$
    and scatters a field $\mathbf{E}_\text{scat}$, which results in a net optical force $\mathbf{F}$.}
    \label{fig:2}
    \end{minipage}
\end{figure}

Once the electromagnetic fields are solved in the discretized finite-element model,
we calculate the resulting forces on the particle using the Maxwell stress tensor (MST) formalism \cite{novotny_principles_2012}.
The basic idea is sketched in \autoref{fig:2}, where a particle scatters an incident electromagnetic field $\mathbf{E}_\text{inc}$ creating the scattered field
$\mathbf{E}_\text{scat}$ and a net force $\mathbf{F}$ that acts on the particle. In the most general case the
time-averaged net force is given by \cite{novotny_principles_2012}:
\begin{equation}\label{eq:MST_force}
    \langle\mathbf{F}\rangle=\int_{\partial
        V}\langle\stackrel{\leftrightarrow}{\mathbf{T}}(\mathbf{r}, t)\rangle \cdot
    \mathbf{n}_{\partial V}(\mathbf{r}) \, \mathrm{d} a
\end{equation}
where $\partial V$ denotes any boundary enclosing the particle and $\mathbf{n}_{\partial V}$
denotes the unitary vector normal to that boundary. In our specific metalens-particle 
problem we select the boundary $\partial V$ as the bounding box of the domain $\Omega_\delta$, which
completely encloses the particle. The stress-tensor is given by \cite{novotny_principles_2012}:
\begin{equation}\label{eq:MST}
    \stackrel{\leftrightarrow}{\mathbf{T}} (\mathbf{r}, t) =\left[\varepsilon_0
        \varepsilon_r \, \mathbf{\mathcal{E}}(\mathbf{r}, t) \otimes
        \mathbf{\mathcal{E}}(\mathbf{r}, t)+\mu_0 \mu_r
        \,\mathbf{\mathcal{H}}(\mathbf{r}, t) \otimes \mathbf{\mathcal{H}}(\mathbf{r},
        t)-\frac{1}{2}\left(\varepsilon_0 \varepsilon_r \mathcal{E}^2(\mathbf{r},
        t)+\mu_0 \mu_r \mathcal{H}^2(\mathbf{r}, t)\right)
        \stackrel{\leftrightarrow}{\mathbf{I}}\right]\,,
\end{equation}
where $\mathbf{\mathcal{E}}(\mathbf{r}, t) = E (\mathbf{r})\, \text{e}^{-\text{i}\omega t}$  and $\mathbf{\mathcal{H}}
    (\mathbf{r}, t) = H (\mathbf{r}) \, \text{e}^{-\text{i}\omega t}$ are the harmonic time-dependent electric and magnetic fields
respectively, $\otimes$ is the outer product, $\varepsilon_0$ is the free-space
dielectric permittivity,  $\varepsilon_r$ is the relative permittivity of the medium surrounding the particle, $\mu_r$ is the
relative magnetic permeability of the medium surrounding the particle, and
$\stackrel{\leftrightarrow}{\mathbf{I}}$ is the identity-tensor. In the case where the surrounding medium is free-space ($\varepsilon_r=\mu_r=1$) and for TE polarization, the time-average of
the MST expression simplifies to:
\begin{equation}
    \langle \stackrel{\leftrightarrow}{\mathbf{T}} (\mathbf{r}, t) \rangle
    = \frac{1}{2} \Re  \begin{bmatrix}
        -\varepsilon_0 E_z^2 /2 +\mu_0\left(H_x^2-H^2/ 2\right) & \varepsilon_0 \mu_0
        H_x H_y^*                                               & 0                                              \\
        \varepsilon_0 \mu_0 H_y H_x^*                           & - \varepsilon_0 E_z^2 /2 +\mu_0\left(H_y^2-H^2
        / 2\right)                                              & 0                                              \\
        0                                                       &
        0                                                       &
        \varepsilon_0 E_z^2 / 2 -\mu_0 H^2 / 2                                                                   \\
    \end{bmatrix}
\end{equation}
where the magnetic field intensity is given by $H^2=H_x\, H_x^* + H_y\, H_y^*$ and $\bullet^*$ denotes the complex conjugate. %Note that since our problem is two-dimensional to calculate 
%the real force on a equivalent three-dimensional particle, we just multiply it with an out-of-plane
%length $L_z$, which for the model to be accurate should be relatively larger than the in-plane dimensions.
%For this problem we choose $L_z=1$ \textmu m.
Note that for particles of dimensions ($w_\delta$, $h_\delta$) much smaller than the wavelength of the electromagnetic field ($w_\delta,\, h_\delta \ll \lambda$), one may apply the dipole approximation. Calculating the forces using \autoref{eq:MST_force} together with the definition of the MST in \autoref{eq:MST} in the dipole approximation, it is possible to derive that a dipole-like particle feels a force proportional to the gradient of the electric field intensity 
$\langle \mathbf{F} \rangle \propto \boldsymbol{\nabla}\left( |E_z(\mathbf{r})|^2\right)$ \cite{novotny_principles_2012, martinez_de_aguirre_jokisch_omnidirectional_2024}; meaning that it is possible to calculate the optical forces without directly including the particle in the electromagnetic simulation. However, in this work we study particles with sizes close to the wavelength ($w_\delta,\, h_\delta \sim \lambda$) and thus the full MST formalism is necessary to describe to forces acting on the particle.

%In literature, optical trapping and manipulation setups involving metamaterials are illuminated
%using focused lasers. In our problem, we model the excitation as an incident plane wave. To
%calculate a realistic electric field amplitude $E_0$ for the excitation given a input laser power
%$P$, we use the expression of the time-averaged intensity for plane-waves in free-space
%$I=(c\varepsilon_0/2)|E|^2$. Given that the intensity is the power per unit surface $P=I/S$, we 
%can recover $E_0\equiv |E| = \sqrt{(2P)/(c\varepsilon_0 S)}$. For the dimension sof our system,
%where $w_\phi=200$ nm and considering a relatively larger out-of-plane length of $1$ \textmu m the 
%amplitude of the electric field for the an input power of $P=1$ W is $E_0\simeq 6 \cdot 10^{7}$ V/m.

To enable the design of the particle and the metalens based on MST force calculations, we introduce a design field, which is defined as $\xi(\mathbf{r})\in[0,1]$. To regularize the optimization problem and avoid pixel-by-pixel variations while introducing a weak sense of geometric length scale \cite{christiansen_compact_2021}, we
apply a filtering and thresholding scheme on the design variables. The thresholding is obtained through a smoothed Heaviside projection \cite{wang_projection_2011}:
\begin{equation}\label{eq:thres}
    \bar{\tilde{\xi}} \equiv \Theta(\tilde{\xi}, \beta, \eta) = \frac{\tanh (\beta \cdot \eta)+\tanh (\beta
        \cdot(\tilde{\xi}-\eta))}{ \tanh (\beta \cdot \eta)+\tanh (\beta
        \cdot(1-\eta))}, \quad \beta \in[1, \infty[, \eta \in[0,1]\,,
\end{equation}
where $\bar{\tilde{\xi}}$ is the physical field, $\Theta(x,\beta,\eta)$ 
is the thresholding function, $\beta$ and $\eta$ control the threshold sharpness and value
respectively, and the filtered design field $\tilde{\xi}$ is obtained through convolution-based filter operation:
\begin{equation}\label{eq:filter}
    \tilde{\xi} (\mathbf{r})= \frac{\sum_{k \in \mathcal{B}_{e, h}}
        w\left(\mathbf{r}-\mathbf{r}_k\right) A_k \xi_k}{\sum_{k \in \mathcal{B}_{e,
                h}} w\left(\mathbf{r}-\mathbf{r}_k\right) A_k}, \quad
    w(\mathbf{r})=\Bigg\{\begin{array}{l}
        r_f-|\mathbf{r}| \quad \forall|\mathbf{r}| \leq r_f \\
        0 \quad \quad \quad \,\, \,
 \forall|\mathbf{r}| > r_f
    \end{array}, \quad r_f \geq 0, \quad \mathbf{r} \in \Omega\\\,,
\end{equation}
where $A_k$ is the area of the $k^\text{th}$ element, $\mathcal{B}_{e,h}$
denotes the $h^\text{th}$ set of finite elements whose center point is within the filter radius
$r_f$ of the $h^\text{th}$ element. Furthermore, in this filtering and thresholding scheme we apply a continuation on $\beta$, which is
essential to enable the use of gradient-based methods while promoting a final binary design. \\

The physical field is then used to interpolate the
dielectric permittivity between free-space ($\varepsilon_r=1$) and an arbitrary
material with relative permittivity $\varepsilon_{r}$, which can be different for the metalens and the particle  \cite{christiansen_compact_2021}:
\begin{equation}\label{eq:mat_int}
    \varepsilon_r( \bar{\tilde{\xi}}(\mathbf{r}))=1+ \bar{\tilde{\xi}}(\mathbf{r})\left(\varepsilon_{r}-1\right)-\mathrm{i} \alpha  \bar{\tilde{\xi}}(\mathbf{r})(1- \bar{\tilde{\xi}}(\mathbf{r})), \quad
    \mathbf{r} \in \Omega
\end{equation}
where $\alpha$ is a problem-dependent artificial attenuation factor. The non-physical
imaginary attenuation factor discourages intermediate values of $\xi$ by introducing
attenuation, or optical losses, in the electromagnetic problem \cite{jensen_topology_2005}.\\

%The expression introduced in \autoref{eq:MST_force} is designed to be applied
%in binary interfaces, where there is a clear distinction of materials at both
%sides of the interface. Howeverm, this is not the case when using inverse
%design to solve the continuous optimization problem, where there might be
%intermediate values of the dielectric permittivity
%$\varepsilon_r(\overline{\tilde{\xi}})$ as given by the material interpolation
%in \autoref{eq:mat_int}. In this formulation we calculate the force as:
%\begin{equation}\label{eq:MST_force_topopt}
%    \langle\mathbf{F}\rangle=\int_{\Gamma}\langle\stackrel{\leftrightarrow}{\mathbf{T}}(\mathbf{r},
%    t)\rangle \cdot \boldsymbol{\nabla} \bar{\tilde{\xi}}(\mathbf{r}) \mathrm{d} a
%\end{equation}
%where $\Gamma$ are all the element boundaries and $  \boldsymbol{\nabla} \bar{\tilde{\xi}}$ gives a measure of change in the design field
%for diffuse boundaries. For the discretized finite-element system, we
%approximate the integral over the boundary by summing the integral over the
%finite-element edges. This is shown in the zoom-in in \autoref{fig:model},
%where the contributions per edge vary depending on the difference of the design
%field with respect to neighboring elements. Note that for purely binary designs
%where there are no intermediate values of the design fields, one recover the
%expression in \autoref{eq:MST_force}.
Following the MST formalism, using \autoref{eq:MST_force} we calculate the resulting force on the single particle defined by the physical design field. Note that to accurately represent the physics we cannot allow the particle to be composed of disconnected members or sub-particles, 
since these would feel different forces in different directions. Similarly, to motivate three-dimensional realizability of the designs, the lens cannot physically have free floating members that are disconnected to the substrate that supports the metalens, since these would collapse
under gravity. To avoid this, we implement a connectivity constraint, where we model the connectivity problem using an artificial
heat-transfer model \cite{li_structural_2016,christiansen_inverse_2023}.
This requires solving an additional partial differential equation:
\begin{equation}\label{eq:heat}
    \nabla \cdot(-c \, (\tilde{\xi}) \nabla C(\mathbf{r})) = f_S({\tilde{\xi}}), \quad \mathbf{r} \in \Omega \in \mathbb{R}^2,
\end{equation}
where we solve for the connectivity (or artificial temperature) field $C(\mathbf{r})$ for a conductivity coefficient $c$ and an excitation $f_S$
subject to the boundary conditions:
\begin{equation}
C=0 \quad \forall \mathbf{r} \in \Gamma_D \,,
\end{equation}
where $\Gamma_D$ are the boundaries we want to connect our solid features to. For the particle we select the boundary to be located at the center point $\mathbf{r}_\delta$, while for the metalens we select the bottom boundary. The rest of the boundaries $\Gamma_N$ are treated with insulating boundary conditions
\begin{equation}
\mathbf{n} \cdot \nabla C=0 \quad \forall \mathbf{r} \in \Gamma_N \,.
\end{equation}
The conductivity coefficient and 
the excitation are given by a material interpolation dependent on the filtered design variables:
\begin{equation}
    \begin{aligned}
    f_S(\tilde{\xi}) & =S_0+\left(S_1-S_0\right) \Theta(\tilde{\xi}, \beta, \eta_C)\, \\
    c(\tilde{\xi}) & =c_0+\left(c_1-c_0\right) \Theta(\tilde{\xi}, \beta, \eta_C)
    \end{aligned}
    \end{equation}
where the constants $c_0$ and $c_1$ denote the artificial conductivity of the material and background respectively, the constants $S_0$ and $S_1$ denote the artificial heat generated by the background and the material
respectively, and where $\eta_C$ controls the threshold value for the conductivity problem.
We have chosen $\eta_C=0.7$, which helps ensure a connected device by acting on an
eroded version of the filtered design field $\tilde{\xi}$. 
Using this material interpolation to solve the heat problem, we calculate a constraint, which acts as a measure for
the total artificial temperature in the simulation domains:
\begin{equation}
    10\log_{10}\left(\frac{\int_{\Omega_i} C(\mathbf{r}) \mathrm{d}\Omega}{\int_{\Omega_i} \mathrm{d}\Omega}\right) \leq \epsilon_C \,, \quad i \in \{\phi, \delta\},\
\end{equation} 
where $\epsilon_C$ is a sufficiently small constant, introduced to enable continuation and alleviate numerical issues when evaluating the constraint. \\

Under these constraints we optimize a FOM that is a spatial projection of the optical
time-averaged forces:
\begin{equation}\label{eq:FOM}
\text{FOM} \equiv \Phi(\langle \mathbf{F}\rangle) = \langle \mathbf{F}\rangle \cdot \mathbf{n}\,,
\end{equation}
where $\mathbf{n}$ is a freely selectable unit vector that can be tailored based on the target application. Note that it is also possible to define more advanced FOMs using optical forces. For instance, it is possible to  target torques by considering the cross product of the distance from a center with respect to the integrand, or to modify the integration limits to target forces  for different sub-regions of the particle.\\

Using the definition of the FOM in \autoref{eq:FOM} and the constraints, we can write out the optimization problem as:
\begin{equation}
    \begin{aligned}\label{eq:opt}
     \max _{\xi}&: \Phi(\mathbf{E}_z) \\
     \text { s.t.}\, &:  \mathbf{S}\left(\varepsilon_r\right) \mathbf{E}_z=\mathbf{f} \text {, } \\
    & :  \mathbf{S}_C\left(c\right) \mathbf{C} = \mathbf{f}_S,\\
    & : 10\log_{10}\left(\frac{\int_{\Omega_i} C(\mathbf{r}) \mathrm{d}\Omega}{\int_{\Omega_i} \mathrm{d}\Omega}\right) \leq \epsilon_C \,,  \quad i \in \{\phi, \delta\}, \\
    %& : \varepsilon_{r, j}=1+\bar{\tilde{\xi}}_j\left(\varepsilon_{r, m}-1\right)-\mathrm{i} \bar{\tilde{\xi}}_j\left(1-\bar{\tilde{\xi}}_j\right) \quad \forall j \in\left\{1,2, \ldots, \mathcal{N}_e\right\} \text {, } \\
    & : 0<\xi_j<1 \quad \forall j \in\left\{1,2, \ldots, \mathcal{N}_D\right\} \text {, } \\
    & : \xi=0 \quad \forall \mathbf{r} \in \Omega /\left\{\Omega_\phi, \Omega_\delta, \Omega_S\right\} \vee \xi=1 \quad \forall \mathbf{r} \in \Omega_S \\
    &
    \end{aligned}
    \end{equation}
where $\mathbf{S}$ is the discretized system matrix which encodes the operators in \autoref{eq:helmholtz},
$\mathbf{E}_z$ and $\mathbf{f}$ are the vectors containing the nodal degrees-of-freedom for the electric 
field and the forcing term in the electromagnetic problem respectively, $\mathbf{S}_C$ is the discretized system matrix which encodes the operators in \autoref{eq:heat}, $\mathbf{C}$ and $\mathbf{f}_C$ are the vectors containing the nodal degrees-of-freedom for the conductivity 
field and the forcing term in the artificial heat problem respectively,  and $\mathcal{N}_D$ denotes the number of design elements. The optimization problem is solved using the method of moving asymptotes (MMA)
 as the optimizer \cite{svanberg_method_1987}.

\section{Adjoint sensitivity analysis}
To enable the use of efficient gradient-based optimizers (MMA) in our topology optimization framework, we need to know how the FOM changes when perturbing the design variables. 
This information is given by the gradients ($\partial \Phi / \partial \xi$) or sensitivities, 
which can be calculated by solving an extra linear system of equations, known as the adjoint problem. To calculate the 
sensitivites we add a zero to the FOM twice by using the discretized Maxwell's equations:
\begin{equation}
    \tilde{\Phi} = \Phi + \boldsymbol{\lambda}^\top (\mathbf{S}(\varepsilon)
    \mathbf{E}_z-\mathbf{f}) + \boldsymbol{\lambda}^\dagger (\mathbf{S}^*(\varepsilon)
    \mathbf{E}_z^*-\mathbf{f}^*)\,,
\end{equation}
where $\boldsymbol{\lambda}$ is a vector of design field independent Lagrange multipliers and $\bullet^\dagger$ denotes the conjugate transpose. We calculate the sensitivity with respect to the physical design field:
\begin{equation}
    \frac{\partial\tilde{\Phi}}{\partial\bar{\tilde{\xi}}} =
    \frac{\partial\Phi}{\partial \mathbf{E}_z}\frac{\partial \mathbf{E}_z}{\partial\bar{\tilde{\xi}}} + \boldsymbol{\lambda}^\top
    \left(\frac{\partial \mathbf{S}(\varepsilon)}{\partial\bar{\tilde{\xi}}} \mathbf{E}_z+
    \mathbf{S}(\varepsilon) \frac{\partial \mathbf{E}_z}{\partial\bar{\tilde{\xi}}}\right) +
    \boldsymbol{\lambda}^\dagger
    \left(\frac{\partial \mathbf{S}^*(\varepsilon)}{\partial\bar{\tilde{\xi}}} \mathbf{E}_z^*+
    \mathbf{S}^*(\varepsilon) \frac{\partial \mathbf{E}_z^*}{\partial\bar{\tilde{\xi}}}\right)\,,
\end{equation}
and we group the terms as:
\begin{equation}
    \frac{\partial\tilde{\Phi}}{\partial\bar{\tilde{\xi}}} = 
    \frac{\partial \mathbf{E}_z}{\partial\bar{\tilde{\xi}}} \underbrace{\left( \frac{\partial
        \Phi}{\partial \mathbf{E}_z}+\boldsymbol{\lambda}^\top \mathbf{S}(\varepsilon) \right)}_{(1)} +
    \frac{\partial \mathbf{E}_z^*}{\partial\bar{\tilde{\xi}}} \underbrace{\left( \frac{\partial
        \Phi}{\partial \mathbf{E}_z^*}+\boldsymbol{\lambda}^\dagger \mathbf{S}^*(\varepsilon) \right)}_{(2)}
    + \boldsymbol{\lambda}^\top \frac{\partial \mathbf{S}(\varepsilon)}{\partial\bar{\tilde{\xi}}}
    \mathbf{E}_z + \boldsymbol{\lambda}^\dagger
    \frac{\partial \mathbf{S}^*(\varepsilon)}{\partial\bar{\tilde{\xi}}} \mathbf{E}_z^*\,.
\end{equation}
We then then determine the Lagrange multipliers such that the terms in the two parentheses become zero. This is done
by taking the first term (1) and adding it to the conjugate of the second term (2), 
we can calculate the Lagrange multiplier by solving the linear system of equations defined by the adjoint problem:
\begin{equation} \label{eq:adj_sys}
    \boldsymbol{\lambda}^\top \mathbf{S} = -\frac{1}{2} \left[ \frac{\partial \Phi}{\partial
            \mathbf{E}_z} + \left( \frac{\partial \Phi}{\partial \mathbf{E}_z^*}\right)^*
        \right]^\top\,,
\end{equation}
which yields the sensitivities:
\begin{equation}
    \frac{\partial\tilde{\Phi}}{\partial\bar{\tilde{\xi}}} =  2 \Re\left\{
        \boldsymbol{\lambda}^\top  \frac{\partial \mathbf{S}(\varepsilon)}{\partial\bar{\tilde{\xi}}}
    \mathbf{E}_z \right\}\,.
\end{equation}
Finally, to obtain the sensitivities of the FOM with respect to the design variables, we apply the chain rule:
\begin{equation}
    \frac{\partial\tilde{\Phi}}{\partial \xi} = \frac{\partial\tilde{\Phi}}{\partial\bar{\tilde{\xi}}} \frac{\partial \bar{\tilde{\xi}}}{\partial \tilde{\xi}} \frac{\partial \tilde{\xi}}{\partial \xi}\,,
\end{equation}
which needs the calculation of the sensitivities of the filter operation in \autoref{eq:filter}
and the threshold operation in \autoref{eq:thres}. Finally to evaluate \autoref{eq:adj_sys} we need to compute
%To calculate the sensitivity
%of the with respect to the physical field for a finite element $e$ yields :
%\begin{equation}
%    \frac{\partial \Phi}{\partial\bar{\tilde{\xi}}_e} = \sum^3_i \frac{\partial f}{\partial \langle F_i \rangle}  \left(\langle\stackrel{\leftrightarrow}{\mathbf{T}}(\mathbf{r},
%    t)\rangle \cdot \mathbf{n}_i \right)
%    \cdot \boldsymbol{\nabla}\mathbf{N}_e \mathrm{d}a  \,
%\end{equation}
%where $i$ denotes the spatial direction, $\mathbf{N}_e$ are the finite-element shape-functions and $\mathrm{d}a$ corresponds to the finite-elements
%discretization length. 
the sensitivity with respect to the electric field which is given by:
\begin{equation}
    \frac{\partial \Phi}{\partial \mathbf{E}_z}= \left(\int_{\partial V}
  \frac{\partial \langle\stackrel{\leftrightarrow}{\mathbf{T}}(\mathbf{r},
 t) \rangle}{\partial \mathbf{E}_z}   \cdot \mathbf{n}_{\partial V} \, \mathrm{d} a\right)\cdot \mathbf{n}\,.
\end{equation}
The sensitivity of the MST with respect to the electric field solution is:
\begin{equation}
    \frac{\partial \langle\stackrel{\leftrightarrow}{\mathbf{T}}(\mathbf{r},
        t)\rangle}{\partial \mathbf{E}_z}
    = \frac{1}{4}\Re \begin{bmatrix}
        -\varepsilon_0 \nu_{z} +\mu_0(\gamma_{x}-\gamma_{y}) & 2\,\varepsilon_0 \mu_0
        \gamma_{xy}                                     & 0                                               \\
        2\, \varepsilon_0 \mu_0 \gamma_{yx}             & -\varepsilon_0 \nu_{z}
        +\mu_0(\gamma_{y}-\gamma_{x})                         & 0                                               \\
        0                                               &
        0                                               & \varepsilon_0 \nu_{z} -\mu_0(\gamma_{y}+ \gamma_{x}) \\
    \end{bmatrix}\,,
\end{equation}
where the individual terms in the tensor are calculated using the relationships in \autoref{eq:mag_fields}:
\begin{align}
    \nu_{z} &\equiv \frac{\partial (\mathbf{E}_z\cdot \mathbf{E}_z^*)}{\partial \mathbf{E}_z} = \mathbf{E}_z^*\,, \\
    \gamma_{x} &\equiv \frac{\partial (\mathbf{H}_x\cdot \mathbf{H}_x^*)}{\partial \mathbf{E}_z} = -\frac{2\, \Im \{ \mathbf{H}_x \}}{\mu_0 \omega}  \cdot \frac{\partial \mathbf{N}}{\partial y}\,, \\
    \gamma_{y} &\equiv \frac{\partial (\mathbf{H}_y\cdot \mathbf{H}_y^*)}{\partial \mathbf{E}_z} = \frac{2\, \Im \{ \mathbf{H}_{y} \}}{\mu_0 \omega}  \cdot \frac{\partial \mathbf{N}}{\partial x}\,, \\
    \gamma_{xy} &\equiv \frac{\partial (\mathbf{H}_x\cdot \mathbf{H}_y^*)}{\partial \mathbf{E}_z} = \frac{\mathrm{i}}{\mu_0 \omega} \left( \frac{\partial \mathbf{N}}{\partial y} \cdot \mathbf{H}^*_{y} + \mathbf{H}_{x} \cdot \frac{\partial \mathbf{N}}{\partial x}\right)\,, \\
    \gamma_{yx} &\equiv \frac{\partial (\mathbf{H}_y\cdot \mathbf{H}_x^*)}{\partial \mathbf{E}_z} = -\frac{\mathrm{i}}{\mu_0 \omega}\left( \frac{\partial \mathbf{N}}{\partial x} \cdot \mathbf{H}^*_{x} + \mathbf{H}_{y} \cdot \frac{\partial \mathbf{N}}{\partial y}\right)\,,
\end{align}
where $\Im$ denotes the imaginary part of a complex field, and $\mathbf{N}$ represent the finite-element interpolation functions. Note that similar results may be derived for the sensitivity with respect to the conjugate field
 $\mathbf{E}_z^*$.

 \section{Examples}

This section details application examples conducted to investigate optimized free-space particles and metalens-particle systems. As an example case study, we operate in the optical regime, at an arbitrary wavelength of 
$\lambda=700$ nm. The entire simulation domain ($\Omega$) has a total height of $h_\Omega=2.24$ \textmu m and a total width of $w_\Omega=4.24$ \textmu m, so that it can fit several wavelengths. In this simulation domain we choose a square particle design domain ($\Omega_\delta$) centered around $\mathbf{r}_\delta=(x,y)=$ (0 nm, 200 nm) with a size equal to the wavelength $w_\delta=h_\delta=700$ nm, so that the dipole approximation is not valid and the MST formalism is required to accurately describe the system. For the metalens region ($\Omega_\phi$) we fix a height $h_\phi=200$ nm and a width of $w_\phi=4$ \textmu m, and for the substrate we select a height $h_s=200$ nm. The model is then discretized using a structured quadrilateral mesh with 20 nm-sized elements. To ensure the validity of the results in the following sections we have conducted a mesh-convergence study as well as a comparison to a commercial tool (COMSOL) ensuring correct implementation and numerical accuracy, finding that at least 10 elements per wavelength were enough to resolve the physics to sufficient accuracy. For the plane-wave excitation we choose an arbitrary amplitude of $E_0=10^{6}$ V/m. As material parameters, we select a relative permittivity of $\varepsilon_r=4$ for both metalens and particle. Note that in Equation 4 we integrate the forces only in two dimensions, meaning that the result will be given in units of force over distance, since we do not integrate over the out-of-plane length of the particle. In the following sections we consider this normalization to be over an out-of-plane length of 1 \textmu m, and thus forces will be given in units of pN/\textmu m.\\

Lastly, for the inverse design hyper-parameters we choose a homogeneous initial guess of $\xi=0.7$ for the design field, a filter radius of $r_f=120$ nm, a threshold value $\eta=0.5$, an initial artificial attenuation $\alpha=0$ and an initial threshold sharpness $\beta=2.5$. Through a continuation scheme the threshold sharpness and the artificial attenuation 
will systematically be increased in the optimization to obtain a final binary design. Note that since the formulated optimization problem is non-convex, global optimality is not guaranteed, and another choice of hyper-parameters may result in a different optimized design.

\subsection{Shaping the particle: maximizing the force in free-space}

\begin{figure}[ht]
    \centering
    \begin{minipage}{0.9\textwidth}
    \includegraphics*[width=90mm]{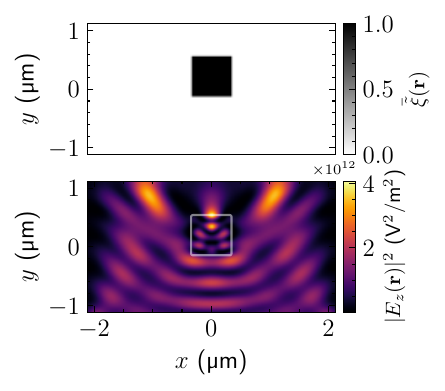}
    \caption{Baseline square particle design and electromagnetic response when illuminated by a plane wave.  On the top the physical density field $\bar{\tilde{\xi}}$ and on the bottom the electric field intensity $|E_z|^2$.}
    \label{fig:cube}
    \end{minipage}
\end{figure}

We first consider the case of a particle floating in free space under the influence of an external plane-wave field excitation, so that there is no metalens structure ($\xi=0\,,\, \forall \mathbf{r} \in \Omega_\phi$). As a baseline to compare the optimized designs in the following sections, we compute the field and the force for a square-like particle where we fix the design variables as: $\xi=1\,,\, \forall \mathbf{r} \in \Omega_\delta$. As shown in the electric field intensity plot in \autoref{fig:cube}, the non-optimal square partly reflects and partly scatters the incoming plane wave, yielding an upward force per micron extrusion of $\langle \mathbf{F}_y \rangle = 2.80$ pN/\textmu m. We also compute a negligible horizontal component of the force per unit length $\langle \mathbf{F}_x \rangle \sim 0$ pN/\textmu m, a result of the axisymmetry of the excitation and the design. Note that these force calculations are normalized by an out-of-plane length of 1 \textmu m. This normalization will also be applied to all other force computations in the following sections.

In this example where a particle is excited by a plane wave, we want to find a particle design that maximizes the vertical component of the force. To do this, we define the FOM as 
$\Phi = \langle \mathbf{F}_y \rangle$, for which we choose a unitary vector in the $y$ direction $\mathbf{n}=\mathbf{n}_y=\binom{0}{1}$ in \autoref{eq:FOM}. We then solve the optimization problem using a maximum of 300 iterations, while also applying a continuation scheme where we increase $\beta^\prime=1.5\,\beta$ and $\alpha^\prime=0.1+\alpha$
when $|\text{FOM}_i-\text{FOM}_{i-1}|\leq 5\cdot 10^{-4}$, where $i$ is the iteration number, and $\beta^\prime$ and $\alpha^\prime$ are the new values of the Heaviside threshold sharpness and artificial attenuation, respectively.\\

\begin{figure}[ht]

    %\hspace*{-1.25cm}
    %\vspace*{0.15cm}
    \centering
    \begin{minipage}{0.9\textwidth}
    \includegraphics*[width=90mm]{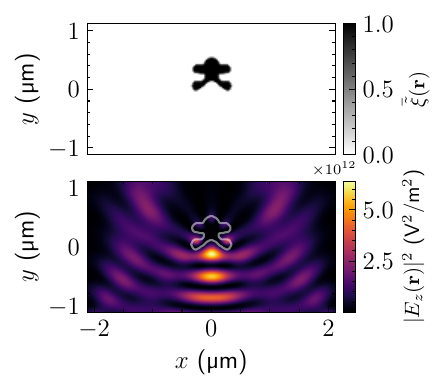}
    \caption{Optimized particle design for the maximization of the 
     $\langle \mathbf{F}_y \rangle $ component of the time-averaged optical force and electromagnetic response when illuminated by a plane wave. On the top the physical density field $\bar{\tilde{\xi}}$ and on the bottom the electric field intensity $|E_z|^2$.}
    \label{fig:repulsive_free}
    \end{minipage}

\end{figure}

We show the optimized design of the particle and the corresponding electric field intensity in \autoref{fig:repulsive_free}. In contrast to the baseline square particle, the optimized design is seen to now reflect most of the plane wave, given that there is almost no transmission above the particle. The optimizer finds a structure that works as a localized Bragg mirror, where the geometric features in the $y$ axis have distances $\sim \lambda/2$. Interestingly, this can be understood with the well known result
of the \textit{radiation pressure} acting on an infinite interface when illuminated by a monochromatic plane wave at normal incidence \cite{novotny_principles_2012}:
\begin{equation}\label{eq:rad_pres}
P_\text{rad} = \langle\stackrel{\leftrightarrow}{\mathbf{T}}(\mathbf{r},
        t)\rangle \cdot \mathbf{n}_y=\frac{\varepsilon_0}{2} E_0^2\left[1+|r|^2\right]\,,
\end{equation}
where $|r|^2\in [0,1] $ is the reflectance of the interface. To maximize the force in the positive $y$ direction one needs to optimize the reflectance of the finite-sized particle. In our case, the optimizer finds a particle geometry that maximizes the reflectance, yielding a total upward force per unit length of $\langle \mathbf{F}_y \rangle = 8.61$ pN/\textmu m. Indeed, this results in a force enhancement factor of $\sim 3$ for the optimized structure compared to the baseline square particle.  Note that both the optimized and baseline design are limited in reflection by the relatively low relative permittivity. As the problem under consideration exhibits translational symmetry in the $y$-direction by computing the forces for different particle positions, we confirm that the force is always positive and constant, meaning that this particle design can be constantly accelerated upwards in a plane wave field. In  essence, the particle design harnesses radiation pressure, similar to solar sail structures \cite{novotny_principles_2012, Davoyan:21} and  can be accelerated under normal plane-wave illumination. Thus, through dimension and material  modifications, this optimization problem could be translated to the design of particle accelerators or solar sail systems \cite{Davoyan:21}.\\

Following the motivation behind \autoref{eq:rad_pres}, we can compare the radiation and the force experienced by an infinite interface under normal plane-wave illumination to our optimized design. A perfectly reflecting ($|r|^2=1$) interface would experience a radiation pressure $P_\text{rad}=\varepsilon_0E_0^2= 8.85 $ pN/\textmu m$^2$. However, our particle  is not an infinite interface but a finite square particle, which scatters light and has corners that cause diffraction effects. Thus, we consider a perfectly reflective square particle by imposing perfect electric conductor boundary conditions  on the particle boundary ($E_z=0, \,\mathbf{r} \in \partial V$). With this new boundary conditions and operating  in the ray optics regime ($w_\delta,\, h_\delta \gg \lambda$) the square particle should be well approximated by an infinite interface. In this case, we find a radiation pressure $P_\text{rad}= 7.60 $ pN/\textmu m$^2$, which updates the previous result by accounting for scattering and diffraction effects. Notably, the optimized particle in \autoref{fig:repulsive_free} yields a radiation pressure of $P_\text{rad}= 12.30 $ pN/\textmu m$^2$, which is larger than the value for a perfectly reflecting square particle and the infinite interface. This can be explained by integrating the radiation pressure over the particle width ($w_\delta$) and comparing the forces, where the force for the optimized particle is larger than for a perfectly reflecting interface $\langle \mathbf{F}_y \rangle = 6.2$ pN/\textmu m and the perfectly reflecting finite particle $\langle \mathbf{F}_y \rangle = 5.32$ pN/\textmu m. The optimized particle achieves a larger force by utilizing the scattering in the structure to create a larger scattering cross-section that is able to reflect a larger fraction of the incoming plane wave, resulting in a larger radiation pressure.\\

Nevertheless, given that the particle is smaller than the simulation domain, there will be a fraction of the plane wave not seen by the particle. This difference can be seen when we calculate the force over a perfectly reflecting interface with the width of the simulation domain ($w_\Omega=4$ \textmu m). In this case, the incoming plane-wave will be totally reflected, resulting in maximal momentum exchange between the plane-wave and the interface in the simulation domain. We calculate the force for a perfectly reflecting surface across the width of the simulation domain by using \autoref{eq:rad_pres}, and find a total upward force of  $\langle \mathbf{F}_y \rangle = 37.54$ pN/\textmu m, which is $\sim 4$ times larger than the force obtained for our optimized design, and $\sim 13$ times larger than the baseline. In the following sections we will show how (nearly) all energy/momentum that is available in the simulation domain may be harnessed by the use of a metalens structure, hereby approaching the reference value for small particles relative to the model domain width.\\

\begin{figure}[ht]
    \centering
    \begin{minipage}[b]{0.48\textwidth}
        %\hspace*{-1.25cm}
        %\vspace*{0.15cm}
        \centering
        \includegraphics*[width=70mm]{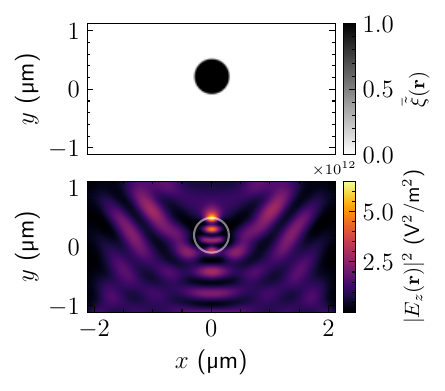}
        \caption*{(a) Baseline circular particle.}
    \end{minipage}
    \hfill
    \begin{minipage}[b]{0.48\textwidth}
        %\hspace*{-1.75cm}
        \centering
        \includegraphics*[width=70mm]{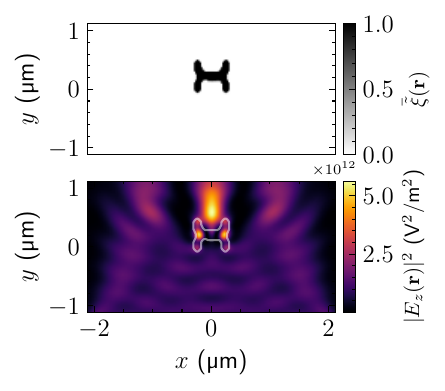}
        \caption*{(b) Optimized attractive particle.}
    \end{minipage}
    \begin{minipage}{1\textwidth}
    \caption{Reference and inverse designed particles, for an attractive particle with gain, where the $\langle \mathbf{F}_y \rangle $ component of the time-averaged optical force is minimized for plane wave illumination. The physical density fields $\bar{\tilde{\xi}}$ and the electric field intensity $|E_z|^2$ are shown
    for both configurations.}
    \label{fig:p_gain}
    \end{minipage}

\end{figure}

Our approach also allows for considering the minimization of the $y$ component of the force for a particle illuminated by a plane wave. However, for a passive material (no gain) illuminated by a single plane-wave excitation with a positive wave vector $\mathbf{k}=k\cdot \mathbf{n}_y$, momentum conservation only allows for a positive $\langle \mathbf{F}_y \rangle$ force component \cite{chenOpticalPullingForce2011}. To achieve an attractive force, also known as \textit{pulling force} \cite{chenOpticalPullingForce2011}, one can employ a material with gain \cite{Lu:24_pulling}. As was shown in \cite{Lu:24_pulling}, by exciting a multipole resonance on a sphere with gain, it is possible to obtain a force that attracts the sphere against the propagation direction of the plane wave. We use a two-dimensional projection of the sphere as a baseline, where we consider a two-dimensional circular particle with gain $\varepsilon_r=4+0.2\mathrm{i}$, as shown in \hyperref[fig:p_gain]{Figure 6 (a)}. By systematically increasing the circle diameter we find that the force on the particle first becomes attractive for a circle diameter of $600$ nm, which allows the excitation of higher multipoles of the circle and results in an attractive force of $\langle \mathbf{F}_y \rangle = -0.12$ pN/\textmu m. Using the same inverse design procedure as for the repulsive particle, while now considering the same gain material ($\varepsilon_r=4+0.2\mathrm{i}$) and a design domain bounding the circle ($w_\delta=h_\delta=600$ nm), we minimize the $\langle \mathbf{F}_y \rangle $ component to obtain an attractive force. This optimization gives the compact design in  \hyperref[fig:p_gain]{Figure 6 (b)}, where the gain particle is maximizing the transmission of the incident plane wave. This results in a net momentum increment of light going through the particle,  which along with momentum conservation results in the particle experiencing an attractive force with a value $\langle \mathbf{F}_y \rangle = -1.65$ pN/\textmu m. The optimized design gives an enhancement factor of $\sim 14$, more than an order of magnitude compared to the baseline circular particle. The force magnitude can be further enhanced by increasing the gain, i.e. the imaginary part of the relative permittivity, or by increasing the size of the design domain. Note that in this example we have modified the complex relative permittivity of the material to consider a particle with gain,  but it can also be modified to considered lossy particles and/or media instead.
 
\subsection{Shaping multibody systems: repulsive and attractive particle-metalens pairs}%COMPARE TO INTEGRATION OVER DOMAIN AND CALCULATE REFLECTION COEFFICIENT

In this section we incorporate the design of the metalens into the system, hereby demonstrating the possibility of tailoring multibody systems. We do this by designing an attractive and a repulsive metalens-particle pairs for a passive material with a purely real effective permittivity ($\varepsilon_r=4$). In order to design a repulsive system
we seek to maximize the vertical force component $\langle \mathbf{F}_y \rangle$, while to design an attractive system we seek to minimize it instead. We achieve this by applying the optimization procedure outlined in the preceding section.\\

\begin{figure}[ht]
    \centering
    \begin{minipage}[b]{0.45\textwidth}
        %\hspace*{-1.25cm}
        %\vspace*{0.15cm}
        \centering
        \includegraphics*[width=70mm]{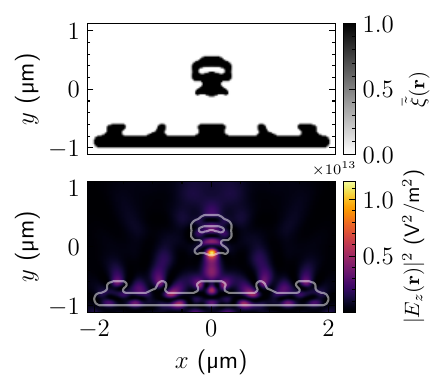}
        \caption*{(a) Attractive metalens-particle system.}
    \end{minipage}
    \hfill
    \begin{minipage}[b]{0.45\textwidth}
        %\hspace*{-1.75cm}
        \centering
        \includegraphics*[width=70mm]{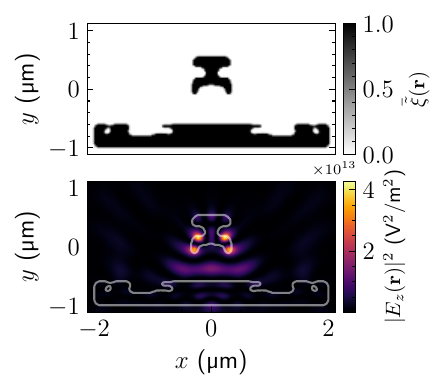}
        \caption*{(b) Repulsive metalens-particle system.}
    \end{minipage}
    \caption{Inverse designed metalens-particle system for the minimization (attractive system) and maximization (repulsive system) of the 
     $\langle \mathbf{F}_y \rangle $ component of the time-averaged optical force. The physical density fields $\bar{\tilde{\xi}}$ and the electric field intensity $|E_z|^2$ are shown
    for both configurations.}
    \label{fig:mp_designs}
\end{figure}

The optimization for the two FOMs yields the results in \autoref{fig:mp_designs}. From the attractive design and its associated electric field intensity distributions we learn that a focusing of the electric field between the particle and the lens results in a net attractive force per micron $\langle \mathbf{F}_y \rangle = -1.64$ pN/\textmu m, which is equivalent to an enhancement factor of $\sim -0.6$ with respect to the baseline design. The strong field enhancement at the bottom of the particle is a combined effect of the light focused by the metalens and the light reflected and refocused by the particle. We interpret that this focusing of the field creates a strong field gradient, resulting in a net force that mainly relies on the optical gradient force, similar to what we 
expect from the dipole approximation picture \cite{martinez_de_aguirre_jokisch_omnidirectional_2024, novotny_principles_2012}. \\

Regarding the optimized repulsive design, the lens helps focus the incident plane wave onto the particle, which similar to the free-space design in the previous section, acts as a reflector. This focused plane-wave and reflector
combination results in a repulsive net force of $\langle \mathbf{F}_y \rangle = 36.28$ pN/\textmu m, which is equivalent to more a than an order of magnitude larger enhancement factor of $\sim 13$ with respect to the baseline and $\sim 4$ times larger than the optimized free-space particle. In fact, this force is close to the reference value $\langle \mathbf{F}_y \rangle = 37.54$ pN/\textmu m  for a perfectly reflecting interface obtained in the previous section, meaning that the metalens is focusing the incoming plane wave onto the particle so that a larger fraction of the incoming energy and momentum available in the simulation domain is exchanged. This results in a much more efficient repulsive device compared to its free-space counterpart. Using \autoref{eq:rad_pres} we find that, in the radiation pressure picture, the optimized device is equivalent to a perfectly reflecting interface with a near unity reflectance of $|r|^2=0.93$. 

\begin{figure}[ht]
    \centering
    \begin{minipage}[b]{0.48\textwidth}
        %\hspace*{-1.25cm}
        %\vspace*{0.15cm}
        \centering
        \includegraphics*[width=70mm]{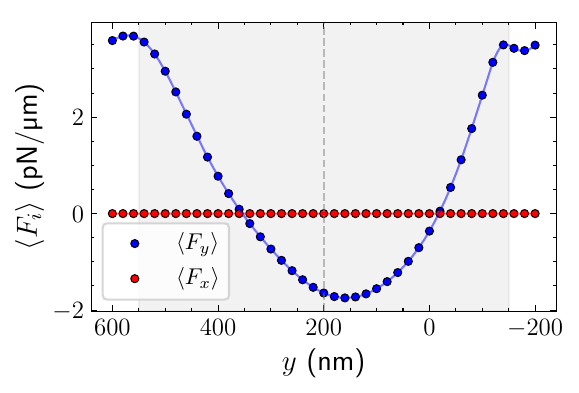}
        \caption*{(a) Attractive metalens-particle system.}
    \end{minipage}
    \hfill
    \begin{minipage}[b]{0.48\textwidth}
        %\hspace*{-1.75cm}
        \centering
        \includegraphics*[width=70mm]{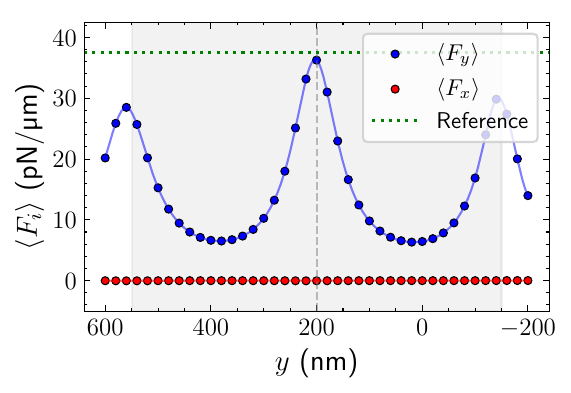}
        \caption*{(b) Repulsive metalens-particle system.}
    \end{minipage}
    \caption{Optical forces as a function of the center of the particle as it is displaced in the $y$ axis for the attractive and repulsive particle-metalens system. The reference value for the radiation pressure force is shown in green for the repulsive system. The $y$ 
    coordinate in the optimization is shown with a dashed line and the particle size is shown as a reference with the shaded region.}
    \label{fig:mp_distance}
\end{figure}

To see how the optimization results translate to other particle positions, we calculate the force acting on the particle when displaced in increments of $20$ nm in the $y$ direction. In \autoref{fig:mp_distance} we show the change on the force components as the particle center is displaced along the $y$ axis. From the results for the attractive particle-metalens system we observe that the force acting upon the particle only remains negative for a distance of around $400$ nm and becomes zero as the particle approaches the center of the simulation domain. If the particle gets too far away from, or to close to the focal spot created by the metalens, it will not be able to feel the field gradient anymore, eliminating the attractive force effect. Interestingly, if the particle is positioned inside the negative force region it will move towards the center of the simulation domain, and then its center-of-mass will start oscillating around $y\sim 20$ nm, due to the harmonic-like force-displacement curve in that region. Regarding the repulsive particle-metalens system we observe how the optical force is always positive, meaning that, similar to the free-space design, it will always get repelled. Moreover, the force is maximal and close to the reference value of the perfectly reflecting interface with the width of the simulation domain, as pointed out previously. Additionally, there are two more local maxima (and minima) separated by a distance $\sim \lambda/2$, hinting at the fact that the particle might be lying at an antinode (node) of the focused plane-wave field, which results in maximal (minimal) positive force. 

\subsection{Towards optimized metalens-based optical trapping: maximizing restoring forces}

The presented framework can also be used to design of effective optical trapping
systems, where the optical forces pull the particles into a stable trapping site. In order to design such an optical system with this formalism, we need to account for the force and field configuration when the particle is displaced
in different directions. A naive formulation for the two-dimensional problem is to account for four
different particle positions, when the particle is slightly displaced ($\Delta r = 80$ nm) in the vertical and horizontal 
axes in both the positive and negative directions. The optimized structure should then exert a maximum
restorative force towards the center position $\mathbf{r}_\delta = (0$ nm$, 200$ nm$)$. To design such a system, we solve four physical
problems for the four different particle positions, and use the \textit{minmax} formulation
for the optimization problem; for an example of such a formulation see \cite{jokisch_topology_2024}. Recasting the non-differentiable
formulation into a so-called bound formulation where the four FOMs are treated as constraints, the optimization problem
is restated as
\begin{equation}
    \begin{aligned}
     \min _{\xi}&: z\\
    \text { s.t.}\,&:  \mathbf{S}_i \left(\varepsilon_r\right) \mathbf{E}_{z,i}=\mathbf{f} , \quad i\in \{\uparrow, \downarrow, \leftarrow, \rightarrow\} \text {, } \\
    & :  \mathbf{S}_C\left(c\right) \mathbf{C} = \mathbf{f}_S,\\
    & : \Phi_{i}(\mathbf{E}_z)-z \leq 0, \quad i\in \{\uparrow, \downarrow, \leftarrow, \rightarrow\}\\
    & : 10\log_{10}\left(\frac{\int_{\Omega_i} C(\mathbf{r}) \mathrm{d}\Omega}{\int_{\Omega_i} \mathrm{d}\Omega}\right) \leq \epsilon_C \,,  \quad i \in \{\phi, \delta\}, \\
    %& : \varepsilon_{r, j}=1+\bar{\tilde{\xi}}_j\left(\varepsilon_{r, m}-1\right)-\mathrm{i} \bar{\tilde{\xi}}_j\left(1-\bar{\tilde{\xi}}_j\right) \quad \forall j \in\left\{1,2, \ldots, \mathcal{N}_e\right\} \text {, } \\
    & : 0<\xi_j<1 \quad \forall j \in\left\{1,2, \ldots, \mathcal{N}_D\right\} \text {, } \\
    & : \xi=0 \quad \forall \mathbf{r} \in \Omega /\left\{\Omega_\phi, \Omega_\delta, \Omega_S\right\} \vee \xi=1 \quad \forall \mathbf{r} \in \Omega_S \\
    &
    \end{aligned}
\end{equation}
where $z$ is an auxiliary optimization variable, and the arrow subscripts denote the particle 
displacement direction for each of the subproblems. In this optimization problem, when we target the force in the horizontal direction
we select a unitary vector in the $x$ direction $\mathbf{n}=\mathbf{n}_x=\binom{w_i}{0}$ in \autoref{eq:FOM}, whereas when we target the vertical component of the force we select 
$\mathbf{n}=\mathbf{n}_y=\binom{0}{w_i}$ instead, where  $w_i$ determines the sign based on the displacement direction; if the displacement is in the positive direction then $w_i=-1$, and if it is in the negative direction then $w_i=1$. The \textit{minmax} formulation then makes sure to maximize the smallest value of the four restoring forces for every iteration, so that the worst-case scenario force is maximized. Additionally, in this section we modify the previously utilized continuation scheme by running the optimization for 250 iterations, where we increase the threshold sharpness 
and attenuation coefficient every 25 iterations, by again updating the values as $\beta^\prime=1.5\,\beta$ and $\alpha^\prime=0.1+\alpha$.\\

\begin{figure}[ht]
    \centering
    \begin{minipage}{0.9\textwidth}
    \includegraphics*[width=90mm]{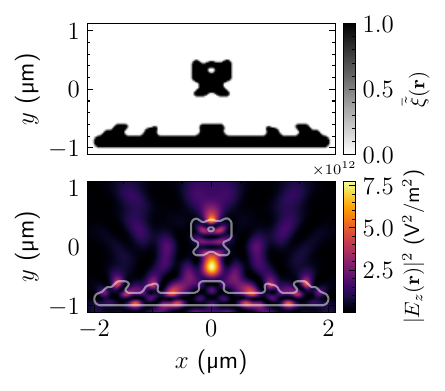}
    \caption{Inverse designed optical trapping metalens-particle system and its electromagnetic response when illuminated by a plane wave. The physical density fields $\bar{\tilde{\xi}}$ and the electric field intensity $|E_z|^2$ are shown
    for the optimized geometry.}
     \label{fig:tweezer}
         \end{minipage}

\end{figure}

This new formulation yields the results in \autoref{fig:tweezer}, and forces per unit length of $\langle F_{y, \uparrow} \rangle = -1.07$ pN/\textmu m, $\langle F_{y, \downarrow} \rangle = 0.78$ pN/\textmu m, $\langle F_{y, \leftarrow} \rangle = 0.90$ pN/\textmu m and $\langle F_{y, \rightarrow} \rangle = -0.90$ pN/\textmu m, which is equivalent to enhancements factor of $\sim \pm 0.3$. This new metalens-particle design is able to create two focusing spots of the electromagnetic fields, one on top of the particle and one on the bottom.
Similar to the attractive metalens-particle design system found in \hyperref[fig:mp_designs]{Figure 2(a)}, we interpret that the design mainly relies on the optical gradient force. There are two spots of strong electric field intensity concentration, one on top of the particle and one below. When
 the particle is displaced in one of the vertical directions its position overlaps with the focusing spot, while the other
spot becomes stronger, making the particle return to the center position. Regarding the horizontal forces, the two focus spots 
balance the horizontal forces of the particle, making the particle return to the center position. This gives the physical effect of an optical trap, ensuring that the particle is effectively drawn back 
to the stable equilibrium point at the geometric center of the particle $\mathbf{r}_\delta$.\\

\begin{figure}[ht]
    \centering
    \begin{minipage}{0.9\textwidth}
    \includegraphics*[width=80mm]{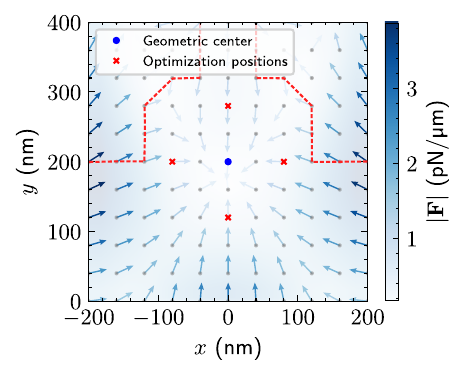}
    \caption{Quiver plot for the two-dimensional forces acting on the optimized particle when evaluated at different spatial positions. The direction of the arrows represents the direction of the force, whereas the colour represents the magnitude. The geometric center of the particle in the optimization $\mathbf{r}_\delta$ is marked in blue, while the positions where the forces were optimized ($\Delta r =80$ nm) are marked in red. The red dashed line delimits the region where the particle is not directed towards the trapping center.}
    \label{fig:quiver}
    \end{minipage}
\end{figure}

Although we have optimized for four different particle positions this still does not guarantee that the particle will return to the center when displaced in an arbitrary direction. However, given that the displacement of the particle is much smaller than the wavelength ($\Delta r \sim \lambda/10$), we expect that in the region delimited by the displacement positions the particle will experience a restoring force toward the center. We validate this in \autoref{fig:quiver}, where we have displaced the particle in increments of $20$ nm from the center position in both horizontal and vertical directions, and have calculated the total force acting on it. Close to the geometric center of the particle (blue dot), the forces point towards a stable trapping site at coordinates $(0$ nm$,180$ nm$)$, yielding a restorative trapping force towards the trapping site for any coordinate in the region delimited by our optimization positions (red crosses). Outside this region there are some domains where the particle is not attracted towards the trapping center, which have been delimited by the red dashed line in \autoref{fig:quiver}. Nevertheless, positioning the particle in the rest of the domain results in restoring forces towards the center, showcasing the robustness of the optimization framework in designing optical trapping systems.

\section{Conclusions}

In this work we have derived a new topology optimization-based inverse design scheme for optical forces that directly incorporates 
Maxwell stress tensor calculations, including expressions for the sensitivities, derived using adjoint sesitivity analysis. It is demonstrated how the scheme can be used to design metalens-particle systems for different optical manipulation
applications; such as, optical trapping and particle acceleration.  Interestingly, we show how, depending on the application, the optimizer not only focuses on
maximizing the gradient forces that were accounted for in other inverse design works \cite{martinez_de_aguirre_jokisch_omnidirectional_2024, nelson_inverse_2024}, but uses the whole spectrum of optical forces included in the MST, such as the radiation pressure force — offering an optimization framework with a more holistic description of optical forces. Moreover, and in contrast to other works \cite{martinez_de_aguirre_jokisch_omnidirectional_2024, nelson_inverse_2024}, the derived framework includes the design of the particle itself, which can further enhance the optical forces. In addition, we benchmark some of the optimized designs against analytical results; such as the radiation pressure force calculated over the width of the simulation domain \cite{novotny_principles_2012}. We also motivate the physical realizability of the devices by considering a connectivity constraint, which
ensures that the optimized devices are connected and that the resulting forces simultaneously affect the whole particle. The optimized geometries and the topology optimization
itself can then be used to design device-particle systems for applications based on optical forces, by carefully modifying 
the FOM function in \autoref{eq:FOM}, making it a good candidate framework to solve more complex optimization problems; such as the design of novel particle accelerators \cite{bar-lev_design_2019}, optical traps \cite{huang_metasurface_2023, xiao_-chip_2023}, optically actuated devices \cite{ivanyi_optically_2024}, active particle systems \cite{bechinger_active_2016}, novel optically-driven particle paths \cite{zemanek_perspective_2019, macdonald_microfluidic_2003, shilkin_directional_2017}, or many-body systems \cite{merrill_many-body_2009, chang_colloquium_2018, bechinger_active_2016}.\\

In future work we foresee the proposed framework being extended in a number of ways. For instance, extending the framework with more advanced FOMs would enable the design of devices that maximize torques enabling the design and optimization of optical rotor devices. Additionally, one could explore alternative phenomena by modifying the plane-wave source to more complex sources, such as non-paraxial beams, by simply modifying the forcing term, $f(\mathbf{r})$, in \autoref{eq:helmholtz}. Moreover, the current scheme could be scaled up to three-dimensional systems by using  appropriate three-dimensional finite elements \cite{jin_finite_2002}, while removing the TE-polarization assumption and solving for the full Maxwell's equations. It is also possible to model the three-dimensional system in two dimensions by including rotational symmetry and obtaining a solution in cylindrical coordinates.

\section*{Acknowledgments}
We would like to thank Jonathan Mirpourian and Benjamin Falkenberg Gøtzsche for useful discussions. We gratefully acknowledge financial support from the
Danish
National Research Foundation through NanoPhoton - Center for Nanophotonics,
grant
number DNRF147.\\

\textbf{Disclosures}
The authors declare no conflicts of interest.

\textbf{Data availability} {The base code underlying the results presented in this paper is available at \texttt{\url{www.topopt.dtu.dk }} and \url{https://github.com/bmdaj/MST_TopOpt}, and the data is available upon reasonable request.

\bibliography{sample}% Produces the bibliography via BibTeX.

\begin{thebibliography}{46}%
\makeatletter
\providecommand \@ifxundefined [1]{%
 \@ifx{#1\undefined}
}%
\providecommand \@ifnum [1]{%
 \ifnum #1\expandafter \@firstoftwo
 \else \expandafter \@secondoftwo
 \fi
}%
\providecommand \@ifx [1]{%
 \ifx #1\expandafter \@firstoftwo
 \else \expandafter \@secondoftwo
 \fi
}%
\providecommand \natexlab [1]{#1}%
\providecommand \enquote  [1]{``#1''}%
\providecommand \bibnamefont  [1]{#1}%
\providecommand \bibfnamefont [1]{#1}%
\providecommand \citenamefont [1]{#1}%
\providecommand \href@noop [0]{\@secondoftwo}%
\providecommand \href [0]{\begingroup \@sanitize@url \@href}%
\providecommand \@href[1]{\@@startlink{#1}\@@href}%
\providecommand \@@href[1]{\endgroup#1\@@endlink}%
\providecommand \@sanitize@url [0]{\catcode `\\12\catcode `\$12\catcode `\&12\catcode `\#12\catcode `\^12\catcode `\_12\catcode `\%12\relax}%
\providecommand \@@startlink[1]{}%
\providecommand \@@endlink[0]{}%
\providecommand \url  [0]{\begingroup\@sanitize@url \@url }%
\providecommand \@url [1]{\endgroup\@href {#1}{\urlprefix }}%
\providecommand \urlprefix  [0]{URL }%
\providecommand \Eprint [0]{\href }%
\providecommand \doibase [0]{https://doi.org/}%
\providecommand \selectlanguage [0]{\@gobble}%
\providecommand \bibinfo  [0]{\@secondoftwo}%
\providecommand \bibfield  [0]{\@secondoftwo}%
\providecommand \translation [1]{[#1]}%
\providecommand \BibitemOpen [0]{}%
\providecommand \bibitemStop [0]{}%
\providecommand \bibitemNoStop [0]{.\EOS\space}%
\providecommand \EOS [0]{\spacefactor3000\relax}%
\providecommand \BibitemShut  [1]{\csname bibitem#1\endcsname}%
\let\auto@bib@innerbib\@empty
%</preamble>
\bibitem [{\citenamefont {Ashkin}(1970)}]{ashkin_acceleration_1970}%
  \BibitemOpen
  \bibfield  {author} {\bibinfo {author} {\bibfnamefont {A.}~\bibnamefont {Ashkin}},\ }\bibfield  {title} {\bibinfo {title} {Acceleration and {Trapping} of {Particles} by {Radiation} {Pressure}},\ }\href {https://doi.org/10.1103/PhysRevLett.24.156} {\bibfield  {journal} {\bibinfo  {journal} {Physical Review Letters}\ }\textbf {\bibinfo {volume} {24}},\ \bibinfo {pages} {156} (\bibinfo {year} {1970})},\ \bibinfo {note} {publisher: American Physical Society}\BibitemShut {NoStop}%
\bibitem [{\citenamefont {Moffitt}\ \emph {et~al.}(2008)\citenamefont {Moffitt}, \citenamefont {Chemla}, \citenamefont {Smith},\ and\ \citenamefont {Bustamante}}]{moffitt_recent_2008}%
  \BibitemOpen
  \bibfield  {author} {\bibinfo {author} {\bibfnamefont {J.~R.}\ \bibnamefont {Moffitt}}, \bibinfo {author} {\bibfnamefont {Y.~R.}\ \bibnamefont {Chemla}}, \bibinfo {author} {\bibfnamefont {S.~B.}\ \bibnamefont {Smith}},\ and\ \bibinfo {author} {\bibfnamefont {C.}~\bibnamefont {Bustamante}},\ }\bibfield  {title} {{\selectlanguage {english}\bibinfo {title} {Recent advances in optical tweezers}},\ }\href {https://doi.org/10.1146/annurev.biochem.77.043007.090225} {\bibfield  {journal} {\bibinfo  {journal} {Annual Review of Biochemistry}\ }\textbf {\bibinfo {volume} {77}},\ \bibinfo {pages} {205} (\bibinfo {year} {2008})}\BibitemShut {NoStop}%
\bibitem [{\citenamefont {Ashkin}(1997)}]{ashkin_optical_1997}%
  \BibitemOpen
  \bibfield  {author} {\bibinfo {author} {\bibfnamefont {A.}~\bibnamefont {Ashkin}},\ }\bibfield  {title} {\bibinfo {title} {Optical trapping and manipulation of neutral particles using lasers},\ }\href {https://doi.org/10.1073/pnas.94.10.4853} {\bibfield  {journal} {\bibinfo  {journal} {Proceedings of the National Academy of Sciences}\ }\textbf {\bibinfo {volume} {94}},\ \bibinfo {pages} {4853} (\bibinfo {year} {1997})},\ \bibinfo {note} {publisher: Proceedings of the National Academy of Sciences}\BibitemShut {NoStop}%
\bibitem [{\citenamefont {Burns}\ \emph {et~al.}(1989)\citenamefont {Burns}, \citenamefont {Fournier},\ and\ \citenamefont {Golovchenko}}]{burns_optical_1989}%
  \BibitemOpen
  \bibfield  {author} {\bibinfo {author} {\bibfnamefont {M.~M.}\ \bibnamefont {Burns}}, \bibinfo {author} {\bibfnamefont {J.-M.}\ \bibnamefont {Fournier}},\ and\ \bibinfo {author} {\bibfnamefont {J.~A.}\ \bibnamefont {Golovchenko}},\ }\bibfield  {title} {\bibinfo {title} {Optical binding},\ }\href {https://doi.org/10.1103/PhysRevLett.63.1233} {\bibfield  {journal} {\bibinfo  {journal} {Physical Review Letters}\ }\textbf {\bibinfo {volume} {63}},\ \bibinfo {pages} {1233} (\bibinfo {year} {1989})},\ \bibinfo {note} {publisher: American Physical Society}\BibitemShut {NoStop}%
\bibitem [{\citenamefont {Dholakia}\ and\ \citenamefont {Zemánek}(2010)}]{dholakia_colloquium_2010}%
  \BibitemOpen
  \bibfield  {author} {\bibinfo {author} {\bibfnamefont {K.}~\bibnamefont {Dholakia}}\ and\ \bibinfo {author} {\bibfnamefont {P.}~\bibnamefont {Zemánek}},\ }\bibfield  {title} {\bibinfo {title} {Colloquium: {Gripped} by light: {Optical} binding},\ }\href {https://doi.org/10.1103/RevModPhys.82.1767} {\bibfield  {journal} {\bibinfo  {journal} {Reviews of Modern Physics}\ }\textbf {\bibinfo {volume} {82}},\ \bibinfo {pages} {1767} (\bibinfo {year} {2010})},\ \bibinfo {note} {publisher: American Physical Society}\BibitemShut {NoStop}%
\bibitem [{\citenamefont {Wang}\ \emph {et~al.}(2005)\citenamefont {Wang}, \citenamefont {Tu}, \citenamefont {Raymond}, \citenamefont {Yang}, \citenamefont {Zhang}, \citenamefont {Hagen}, \citenamefont {Dees}, \citenamefont {Mercer}, \citenamefont {Forster}, \citenamefont {Kariv}, \citenamefont {Marchand},\ and\ \citenamefont {Butler}}]{wang_microfluidic_2005}%
  \BibitemOpen
  \bibfield  {author} {\bibinfo {author} {\bibfnamefont {M.~M.}\ \bibnamefont {Wang}}, \bibinfo {author} {\bibfnamefont {E.}~\bibnamefont {Tu}}, \bibinfo {author} {\bibfnamefont {D.~E.}\ \bibnamefont {Raymond}}, \bibinfo {author} {\bibfnamefont {J.~M.}\ \bibnamefont {Yang}}, \bibinfo {author} {\bibfnamefont {H.}~\bibnamefont {Zhang}}, \bibinfo {author} {\bibfnamefont {N.}~\bibnamefont {Hagen}}, \bibinfo {author} {\bibfnamefont {B.}~\bibnamefont {Dees}}, \bibinfo {author} {\bibfnamefont {E.~M.}\ \bibnamefont {Mercer}}, \bibinfo {author} {\bibfnamefont {A.~H.}\ \bibnamefont {Forster}}, \bibinfo {author} {\bibfnamefont {I.}~\bibnamefont {Kariv}}, \bibinfo {author} {\bibfnamefont {P.~J.}\ \bibnamefont {Marchand}},\ and\ \bibinfo {author} {\bibfnamefont {W.~F.}\ \bibnamefont {Butler}},\ }\bibfield  {title} {{\selectlanguage {english}\bibinfo {title} {Microfluidic sorting of mammalian cells by optical force switching}},\ }\href {https://doi.org/10.1038/nbt1050} {\bibfield  {journal} {\bibinfo  {journal} {Nature
  Biotechnology}\ }\textbf {\bibinfo {volume} {23}},\ \bibinfo {pages} {83} (\bibinfo {year} {2005})},\ \bibinfo {note} {publisher: Nature Publishing Group}\BibitemShut {NoStop}%
\bibitem [{\citenamefont {Almaas}\ and\ \citenamefont {Brevik}(2013)}]{almaas_possible_2013}%
  \BibitemOpen
  \bibfield  {author} {\bibinfo {author} {\bibfnamefont {E.}~\bibnamefont {Almaas}}\ and\ \bibinfo {author} {\bibfnamefont {I.}~\bibnamefont {Brevik}},\ }\bibfield  {title} {\bibinfo {title} {Possible sorting mechanism for microparticles in an evanescent field},\ }\href {https://doi.org/10.1103/PhysRevA.87.063826} {\bibfield  {journal} {\bibinfo  {journal} {Physical Review A}\ }\textbf {\bibinfo {volume} {87}},\ \bibinfo {pages} {063826} (\bibinfo {year} {2013})},\ \bibinfo {note} {publisher: American Physical Society}\BibitemShut {NoStop}%
\bibitem [{\citenamefont {Chang}\ \emph {et~al.}(2009)\citenamefont {Chang}, \citenamefont {Thompson}, \citenamefont {Park}, \citenamefont {Vuletić}, \citenamefont {Zibrov}, \citenamefont {Zoller},\ and\ \citenamefont {Lukin}}]{chang_trapping_2009}%
  \BibitemOpen
  \bibfield  {author} {\bibinfo {author} {\bibfnamefont {D.~E.}\ \bibnamefont {Chang}}, \bibinfo {author} {\bibfnamefont {J.~D.}\ \bibnamefont {Thompson}}, \bibinfo {author} {\bibfnamefont {H.}~\bibnamefont {Park}}, \bibinfo {author} {\bibfnamefont {V.}~\bibnamefont {Vuletić}}, \bibinfo {author} {\bibfnamefont {A.~S.}\ \bibnamefont {Zibrov}}, \bibinfo {author} {\bibfnamefont {P.}~\bibnamefont {Zoller}},\ and\ \bibinfo {author} {\bibfnamefont {M.~D.}\ \bibnamefont {Lukin}},\ }\bibfield  {title} {\bibinfo {title} {Trapping and {Manipulation} of {Isolated} {Atoms} {Using} {Nanoscale} {Plasmonic} {Structures}},\ }\href {https://doi.org/10.1103/PhysRevLett.103.123004} {\bibfield  {journal} {\bibinfo  {journal} {Physical Review Letters}\ }\textbf {\bibinfo {volume} {103}},\ \bibinfo {pages} {123004} (\bibinfo {year} {2009})},\ \bibinfo {note} {publisher: American Physical Society}\BibitemShut {NoStop}%
\bibitem [{\citenamefont {Bustamante}\ \emph {et~al.}(2021)\citenamefont {Bustamante}, \citenamefont {Chemla}, \citenamefont {Liu},\ and\ \citenamefont {Wang}}]{bustamante_optical_2021}%
  \BibitemOpen
  \bibfield  {author} {\bibinfo {author} {\bibfnamefont {C.~J.}\ \bibnamefont {Bustamante}}, \bibinfo {author} {\bibfnamefont {Y.~R.}\ \bibnamefont {Chemla}}, \bibinfo {author} {\bibfnamefont {S.}~\bibnamefont {Liu}},\ and\ \bibinfo {author} {\bibfnamefont {M.~D.}\ \bibnamefont {Wang}},\ }\bibfield  {title} {{\selectlanguage {english}\bibinfo {title} {Optical tweezers in single-molecule biophysics}},\ }\href {https://doi.org/10.1038/s43586-021-00021-6} {\bibfield  {journal} {\bibinfo  {journal} {Nature Reviews Methods Primers}\ }\textbf {\bibinfo {volume} {1}},\ \bibinfo {pages} {1} (\bibinfo {year} {2021})},\ \bibinfo {note} {publisher: Nature Publishing Group}\BibitemShut {NoStop}%
\bibitem [{\citenamefont {Li}\ \emph {et~al.}(2006)\citenamefont {Li}, \citenamefont {Zhou}, \citenamefont {Browne},\ and\ \citenamefont {Klenerman}}]{li_evidence_2006}%
  \BibitemOpen
  \bibfield  {author} {\bibinfo {author} {\bibfnamefont {H.}~\bibnamefont {Li}}, \bibinfo {author} {\bibfnamefont {D.}~\bibnamefont {Zhou}}, \bibinfo {author} {\bibfnamefont {H.}~\bibnamefont {Browne}},\ and\ \bibinfo {author} {\bibfnamefont {D.}~\bibnamefont {Klenerman}},\ }\bibfield  {title} {\bibinfo {title} {Evidence for {Resonance} {Optical} {Trapping} of {Individual} {Fluorophore}-{Labeled} {Antibodies} {Using} {Single} {Molecule} {Fluorescence} {Spectroscopy}},\ }\href {https://doi.org/10.1021/ja056997t} {\bibfield  {journal} {\bibinfo  {journal} {Journal of the American Chemical Society}\ }\textbf {\bibinfo {volume} {128}},\ \bibinfo {pages} {5711} (\bibinfo {year} {2006})},\ \bibinfo {note} {publisher: American Chemical Society}\BibitemShut {NoStop}%
\bibitem [{\citenamefont {Ashkin}(1992)}]{ashkin_forces_1992}%
  \BibitemOpen
  \bibfield  {author} {\bibinfo {author} {\bibfnamefont {A.}~\bibnamefont {Ashkin}},\ }\bibfield  {title} {\bibinfo {title} {Forces of a single-beam gradient laser trap on a dielectric sphere in the ray optics regime},\ }\href {https://doi.org/10.1016/S0006-3495(92)81860-X} {\bibfield  {journal} {\bibinfo  {journal} {Biophysical Journal}\ }\textbf {\bibinfo {volume} {61}},\ \bibinfo {pages} {569} (\bibinfo {year} {1992})}\BibitemShut {NoStop}%
\bibitem [{\citenamefont {Xiao}\ \emph {et~al.}(2023)\citenamefont {Xiao}, \citenamefont {Plaskocinski}, \citenamefont {Biabanifard}, \citenamefont {Persheyev},\ and\ \citenamefont {Di~Falco}}]{xiao_-chip_2023}%
  \BibitemOpen
  \bibfield  {author} {\bibinfo {author} {\bibfnamefont {J.}~\bibnamefont {Xiao}}, \bibinfo {author} {\bibfnamefont {T.}~\bibnamefont {Plaskocinski}}, \bibinfo {author} {\bibfnamefont {M.}~\bibnamefont {Biabanifard}}, \bibinfo {author} {\bibfnamefont {S.}~\bibnamefont {Persheyev}},\ and\ \bibinfo {author} {\bibfnamefont {A.}~\bibnamefont {Di~Falco}},\ }\bibfield  {title} {\bibinfo {title} {On-{Chip} {Optical} {Trapping} with {High} {NA} {Metasurfaces}},\ }\href {https://doi.org/10.1021/acsphotonics.2c01986} {\bibfield  {journal} {\bibinfo  {journal} {ACS Photonics}\ }\textbf {\bibinfo {volume} {10}},\ \bibinfo {pages} {1341} (\bibinfo {year} {2023})},\ \bibinfo {note} {publisher: American Chemical Society}\BibitemShut {NoStop}%
\bibitem [{\citenamefont {Brooks}\ \emph {et~al.}(2019)\citenamefont {Brooks}, \citenamefont {Tasinkevych}, \citenamefont {Sabrina}, \citenamefont {Velegol}, \citenamefont {Sen},\ and\ \citenamefont {Bishop}}]{brooks_shape-directed_2019}%
  \BibitemOpen
  \bibfield  {author} {\bibinfo {author} {\bibfnamefont {A.~M.}\ \bibnamefont {Brooks}}, \bibinfo {author} {\bibfnamefont {M.}~\bibnamefont {Tasinkevych}}, \bibinfo {author} {\bibfnamefont {S.}~\bibnamefont {Sabrina}}, \bibinfo {author} {\bibfnamefont {D.}~\bibnamefont {Velegol}}, \bibinfo {author} {\bibfnamefont {A.}~\bibnamefont {Sen}},\ and\ \bibinfo {author} {\bibfnamefont {K.~J.~M.}\ \bibnamefont {Bishop}},\ }\bibfield  {title} {{\selectlanguage {english}\bibinfo {title} {Shape-directed rotation of homogeneous micromotors via catalytic self-electrophoresis}},\ }\href {https://doi.org/10.1038/s41467-019-08423-7} {\bibfield  {journal} {\bibinfo  {journal} {Nature Communications}\ }\textbf {\bibinfo {volume} {10}},\ \bibinfo {pages} {495} (\bibinfo {year} {2019})},\ \bibinfo {note} {publisher: Nature Publishing Group}\BibitemShut {NoStop}%
\bibitem [{\citenamefont {Lee}\ \emph {et~al.}(2017)\citenamefont {Lee}, \citenamefont {Miller}, \citenamefont {Reid}, \citenamefont {Johnson},\ and\ \citenamefont {Fang}}]{Lee:17}%
  \BibitemOpen
  \bibfield  {author} {\bibinfo {author} {\bibfnamefont {Y.~E.}\ \bibnamefont {Lee}}, \bibinfo {author} {\bibfnamefont {O.~D.}\ \bibnamefont {Miller}}, \bibinfo {author} {\bibfnamefont {M.~T.~H.}\ \bibnamefont {Reid}}, \bibinfo {author} {\bibfnamefont {S.~G.}\ \bibnamefont {Johnson}},\ and\ \bibinfo {author} {\bibfnamefont {N.~X.}\ \bibnamefont {Fang}},\ }\bibfield  {title} {\bibinfo {title} {Computational inverse design of non-intuitive illumination patterns to maximize optical force or torque},\ }\href {https://doi.org/10.1364/OE.25.006757} {\bibfield  {journal} {\bibinfo  {journal} {Optics Express}\ }\textbf {\bibinfo {volume} {25}},\ \bibinfo {pages} {6757} (\bibinfo {year} {2017})},\ \bibinfo {note} {publisher: Optica Publishing Group}\BibitemShut {NoStop}%
\bibitem [{\citenamefont {Nelson}\ \emph {et~al.}(2024)\citenamefont {Nelson}, \citenamefont {Kim},\ and\ \citenamefont {Crozier}}]{nelson_inverse_2024}%
  \BibitemOpen
  \bibfield  {author} {\bibinfo {author} {\bibfnamefont {D.}~\bibnamefont {Nelson}}, \bibinfo {author} {\bibfnamefont {S.}~\bibnamefont {Kim}},\ and\ \bibinfo {author} {\bibfnamefont {K.~B.}\ \bibnamefont {Crozier}},\ }\bibfield  {title} {\bibinfo {title} {Inverse {Design} of {Plasmonic} {Nanotweezers} by {Topology} {Optimization}},\ }\href {https://doi.org/10.1021/acsphotonics.3c01019} {\bibfield  {journal} {\bibinfo  {journal} {ACS Photonics}\ }\textbf {\bibinfo {volume} {11}},\ \bibinfo {pages} {85} (\bibinfo {year} {2024})},\ \bibinfo {note} {publisher: American Chemical Society}\BibitemShut {NoStop}%
\bibitem [{\citenamefont {Martinez~de Aguirre~Jokisch}\ \emph {et~al.}(2024)\citenamefont {Martinez~de Aguirre~Jokisch}, \citenamefont {Gøtzsche}, \citenamefont {Kristensen}, \citenamefont {Wubs}, \citenamefont {Sigmund},\ and\ \citenamefont {Christiansen}}]{martinez_de_aguirre_jokisch_omnidirectional_2024}%
  \BibitemOpen
  \bibfield  {author} {\bibinfo {author} {\bibfnamefont {B.}~\bibnamefont {Martinez~de Aguirre~Jokisch}}, \bibinfo {author} {\bibfnamefont {B.~F.}\ \bibnamefont {Gøtzsche}}, \bibinfo {author} {\bibfnamefont {P.~T.}\ \bibnamefont {Kristensen}}, \bibinfo {author} {\bibfnamefont {M.}~\bibnamefont {Wubs}}, \bibinfo {author} {\bibfnamefont {O.}~\bibnamefont {Sigmund}},\ and\ \bibinfo {author} {\bibfnamefont {R.~E.}\ \bibnamefont {Christiansen}},\ }\bibfield  {title} {\bibinfo {title} {Omnidirectional {Gradient} {Force} {Optical} {Trapping} in {Dielectric} {Nanocavities} by {Inverse} {Design}},\ }\bibfield  {journal} {\bibinfo  {journal} {ACS Photonics}\ }\href {https://doi.org/10.1021/acsphotonics.4c01060} {10.1021/acsphotonics.4c01060} (\bibinfo {year} {2024}),\ \bibinfo {note} {publisher: American Chemical Society}\BibitemShut {NoStop}%
\bibitem [{\citenamefont {Jensen}\ and\ \citenamefont {Sigmund}(2011)}]{jensen_topology_2011}%
  \BibitemOpen
  \bibfield  {author} {\bibinfo {author} {\bibfnamefont {J.}~\bibnamefont {Jensen}}\ and\ \bibinfo {author} {\bibfnamefont {O.}~\bibnamefont {Sigmund}},\ }\bibfield  {title} {\bibinfo {title} {Topology optimization for nano-photonics},\ }\href {https://doi.org/10.1002/lpor.201000014} {\bibfield  {journal} {\bibinfo  {journal} {Laser \& Photonics Reviews}\ }\textbf {\bibinfo {volume} {5}},\ \bibinfo {pages} {308} (\bibinfo {year} {2011})},\ \bibinfo {note} {publisher: John Wiley \& Sons, Ltd}\BibitemShut {NoStop}%
\bibitem [{\citenamefont {Wang}\ \emph {et~al.}(2018)\citenamefont {Wang}, \citenamefont {Christiansen}, \citenamefont {Yu}, \citenamefont {Mørk},\ and\ \citenamefont {Sigmund}}]{wang_maximizing_2018}%
  \BibitemOpen
  \bibfield  {author} {\bibinfo {author} {\bibfnamefont {F.}~\bibnamefont {Wang}}, \bibinfo {author} {\bibfnamefont {R.~E.}\ \bibnamefont {Christiansen}}, \bibinfo {author} {\bibfnamefont {Y.}~\bibnamefont {Yu}}, \bibinfo {author} {\bibfnamefont {J.}~\bibnamefont {Mørk}},\ and\ \bibinfo {author} {\bibfnamefont {O.}~\bibnamefont {Sigmund}},\ }\bibfield  {title} {\bibinfo {title} {Maximizing the quality factor to mode volume ratio for ultra-small photonic crystal cavities},\ }\href {https://doi.org/10.1063/1.5064468} {\bibfield  {journal} {\bibinfo  {journal} {Applied Physics Letters}\ }\textbf {\bibinfo {volume} {113}},\ \bibinfo {pages} {241101} (\bibinfo {year} {2018})}\BibitemShut {NoStop}%
\bibitem [{\citenamefont {Yao}\ \emph {et~al.}(2022)\citenamefont {Yao}, \citenamefont {Verdugo}, \citenamefont {Christiansen},\ and\ \citenamefont {Johnson}}]{yao_trace_2022}%
  \BibitemOpen
  \bibfield  {author} {\bibinfo {author} {\bibfnamefont {W.}~\bibnamefont {Yao}}, \bibinfo {author} {\bibfnamefont {F.}~\bibnamefont {Verdugo}}, \bibinfo {author} {\bibfnamefont {R.~E.}\ \bibnamefont {Christiansen}},\ and\ \bibinfo {author} {\bibfnamefont {S.~G.}\ \bibnamefont {Johnson}},\ }\bibfield  {title} {{\selectlanguage {english}\bibinfo {title} {Trace formulation for photonic inverse design with incoherent sources}},\ }\href {https://doi.org/10.1007/s00158-022-03389-5} {\bibfield  {journal} {\bibinfo  {journal} {Structural and Multidisciplinary Optimization}\ }\textbf {\bibinfo {volume} {65}},\ \bibinfo {pages} {336} (\bibinfo {year} {2022})}\BibitemShut {NoStop}%
\bibitem [{\citenamefont {Albrechtsen}\ \emph {et~al.}(2022)\citenamefont {Albrechtsen}, \citenamefont {Vosoughi~Lahijani}, \citenamefont {Christiansen}, \citenamefont {Nguyen}, \citenamefont {Casses}, \citenamefont {Hansen}, \citenamefont {Stenger}, \citenamefont {Sigmund}, \citenamefont {Jansen}, \citenamefont {Mørk},\ and\ \citenamefont {Stobbe}}]{albrechtsen_nanometer-scale_2022}%
  \BibitemOpen
  \bibfield  {author} {\bibinfo {author} {\bibfnamefont {M.}~\bibnamefont {Albrechtsen}}, \bibinfo {author} {\bibfnamefont {B.}~\bibnamefont {Vosoughi~Lahijani}}, \bibinfo {author} {\bibfnamefont {R.~E.}\ \bibnamefont {Christiansen}}, \bibinfo {author} {\bibfnamefont {V.~T.~H.}\ \bibnamefont {Nguyen}}, \bibinfo {author} {\bibfnamefont {L.~N.}\ \bibnamefont {Casses}}, \bibinfo {author} {\bibfnamefont {S.~E.}\ \bibnamefont {Hansen}}, \bibinfo {author} {\bibfnamefont {N.}~\bibnamefont {Stenger}}, \bibinfo {author} {\bibfnamefont {O.}~\bibnamefont {Sigmund}}, \bibinfo {author} {\bibfnamefont {H.}~\bibnamefont {Jansen}}, \bibinfo {author} {\bibfnamefont {J.}~\bibnamefont {Mørk}},\ and\ \bibinfo {author} {\bibfnamefont {S.}~\bibnamefont {Stobbe}},\ }\bibfield  {title} {{\selectlanguage {english}\bibinfo {title} {Nanometer-scale photon confinement in topology-optimized dielectric cavities}},\ }\href {https://doi.org/10.1038/s41467-022-33874-w} {\bibfield  {journal} {\bibinfo  {journal} {Nature Communications}\
  }\textbf {\bibinfo {volume} {13}},\ \bibinfo {pages} {6281} (\bibinfo {year} {2022})},\ \bibinfo {note} {publisher: Nature Publishing Group}\BibitemShut {NoStop}%
\bibitem [{\citenamefont {Ahn}\ \emph {et~al.}(2022)\citenamefont {Ahn}, \citenamefont {Yang}, \citenamefont {Trivedi}, \citenamefont {White}, \citenamefont {Su}, \citenamefont {Skarda},\ and\ \citenamefont {Vučković}}]{ahn_photonic_2022}%
  \BibitemOpen
  \bibfield  {author} {\bibinfo {author} {\bibfnamefont {G.~H.}\ \bibnamefont {Ahn}}, \bibinfo {author} {\bibfnamefont {K.~Y.}\ \bibnamefont {Yang}}, \bibinfo {author} {\bibfnamefont {R.}~\bibnamefont {Trivedi}}, \bibinfo {author} {\bibfnamefont {A.~D.}\ \bibnamefont {White}}, \bibinfo {author} {\bibfnamefont {L.}~\bibnamefont {Su}}, \bibinfo {author} {\bibfnamefont {J.}~\bibnamefont {Skarda}},\ and\ \bibinfo {author} {\bibfnamefont {J.}~\bibnamefont {Vučković}},\ }\bibfield  {title} {\bibinfo {title} {Photonic {Inverse} {Design} of {On}-{Chip} {Microresonators}},\ }\href {https://doi.org/10.1021/acsphotonics.2c00020} {\bibfield  {journal} {\bibinfo  {journal} {ACS Photonics}\ }\textbf {\bibinfo {volume} {9}},\ \bibinfo {pages} {1875} (\bibinfo {year} {2022})},\ \bibinfo {note} {publisher: American Chemical Society}\BibitemShut {NoStop}%
\bibitem [{\citenamefont {Chung}\ and\ \citenamefont {Miller}(2020)}]{chung_high-na_2020}%
  \BibitemOpen
  \bibfield  {author} {\bibinfo {author} {\bibfnamefont {H.}~\bibnamefont {Chung}}\ and\ \bibinfo {author} {\bibfnamefont {O.~D.}\ \bibnamefont {Miller}},\ }\bibfield  {title} {{\selectlanguage {english}\bibinfo {title} {High-{NA} achromatic metalenses by inverse design}},\ }\href {https://doi.org/10.1364/OE.385440} {\bibfield  {journal} {\bibinfo  {journal} {Optics Express}\ }\textbf {\bibinfo {volume} {28}},\ \bibinfo {pages} {6945} (\bibinfo {year} {2020})},\ \bibinfo {note} {publisher: Optica Publishing Group}\BibitemShut {NoStop}%
\bibitem [{\citenamefont {Christiansen}\ and\ \citenamefont {Sigmund}(2021)}]{christiansen_compact_2021}%
  \BibitemOpen
  \bibfield  {author} {\bibinfo {author} {\bibfnamefont {R.~E.}\ \bibnamefont {Christiansen}}\ and\ \bibinfo {author} {\bibfnamefont {O.}~\bibnamefont {Sigmund}},\ }\bibfield  {title} {{\selectlanguage {english}\bibinfo {title} {Compact 200 line {MATLAB} code for inverse design in photonics by topology optimization: tutorial}},\ }\href {https://doi.org/10.1364/JOSAB.405955} {\bibfield  {journal} {\bibinfo  {journal} {JOSA B}\ }\textbf {\bibinfo {volume} {38}},\ \bibinfo {pages} {510} (\bibinfo {year} {2021})},\ \bibinfo {note} {publisher: Optica Publishing Group}\BibitemShut {NoStop}%
\bibitem [{\citenamefont {Li}\ \emph {et~al.}(2022)\citenamefont {Li}, \citenamefont {Pestourie}, \citenamefont {Park}, \citenamefont {Huang}, \citenamefont {Johnson},\ and\ \citenamefont {Capasso}}]{li_inverse_2022}%
  \BibitemOpen
  \bibfield  {author} {\bibinfo {author} {\bibfnamefont {Z.}~\bibnamefont {Li}}, \bibinfo {author} {\bibfnamefont {R.}~\bibnamefont {Pestourie}}, \bibinfo {author} {\bibfnamefont {J.-S.}\ \bibnamefont {Park}}, \bibinfo {author} {\bibfnamefont {Y.-W.}\ \bibnamefont {Huang}}, \bibinfo {author} {\bibfnamefont {S.~G.}\ \bibnamefont {Johnson}},\ and\ \bibinfo {author} {\bibfnamefont {F.}~\bibnamefont {Capasso}},\ }\bibfield  {title} {{\selectlanguage {english}\bibinfo {title} {Inverse design enables large-scale high-performance meta-optics reshaping virtual reality}},\ }\href {https://doi.org/10.1038/s41467-022-29973-3} {\bibfield  {journal} {\bibinfo  {journal} {Nature Communications}\ }\textbf {\bibinfo {volume} {13}},\ \bibinfo {pages} {2409} (\bibinfo {year} {2022})},\ \bibinfo {note} {publisher: Nature Publishing Group}\BibitemShut {NoStop}%
\bibitem [{\citenamefont {Novotny}\ and\ \citenamefont {Hecht}(2012)}]{novotny_principles_2012}%
  \BibitemOpen
  \bibfield  {author} {\bibinfo {author} {\bibfnamefont {L.}~\bibnamefont {Novotny}}\ and\ \bibinfo {author} {\bibfnamefont {B.}~\bibnamefont {Hecht}},\ }\href {https://doi.org/10.1017/CBO9780511794193} {\emph {\bibinfo {title} {Principles of {Nano}-{Optics}}}},\ \bibinfo {edition} {2nd}\ ed.\ (\bibinfo  {publisher} {Cambridge University Press},\ \bibinfo {address} {Cambridge},\ \bibinfo {year} {2012})\BibitemShut {NoStop}%
\bibitem [{\citenamefont {Glückstad}(2011)}]{gluck}%
  \BibitemOpen
  \bibfield  {author} {\bibinfo {author} {\bibfnamefont {J.}~\bibnamefont {Glückstad}},\ }\bibfield  {title} {\bibinfo {title} {Optical manipulation: Sculpting the object},\ }\href {https://doi.org/10.1038/nphoton.2010.301} {\bibfield  {journal} {\bibinfo  {journal} {Nature Photonics}\ }\textbf {\bibinfo {volume} {5}} (\bibinfo {year} {2011})}\BibitemShut {NoStop}%
\bibitem [{\citenamefont {Rodriguez}\ \emph {et~al.}(2014)\citenamefont {Rodriguez}, \citenamefont {Hui}, \citenamefont {Woolf}, \citenamefont {Johnson}, \citenamefont {Lončar},\ and\ \citenamefont {Capasso}}]{Rodriguez_2014}%
  \BibitemOpen
  \bibfield  {author} {\bibinfo {author} {\bibfnamefont {A.~W.}\ \bibnamefont {Rodriguez}}, \bibinfo {author} {\bibfnamefont {P.}~\bibnamefont {Hui}}, \bibinfo {author} {\bibfnamefont {D.~P.}\ \bibnamefont {Woolf}}, \bibinfo {author} {\bibfnamefont {S.~G.}\ \bibnamefont {Johnson}}, \bibinfo {author} {\bibfnamefont {M.}~\bibnamefont {Lončar}},\ and\ \bibinfo {author} {\bibfnamefont {F.}~\bibnamefont {Capasso}},\ }\bibfield  {title} {\bibinfo {title} {Classical and fluctuation‐induced electromagnetic interactions in micron‐scale systems: designer bonding, antibonding, and casimir forces},\ }\href {https://doi.org/10.1002/andp.201400160} {\bibfield  {journal} {\bibinfo  {journal} {Annalen der Physik}\ }\textbf {\bibinfo {volume} {527}},\ \bibinfo {pages} {45–80} (\bibinfo {year} {2014})}\BibitemShut {NoStop}%
\bibitem [{\citenamefont {Huang}\ \emph {et~al.}(2023)\citenamefont {Huang}, \citenamefont {Yuan}, \citenamefont {Holman}, \citenamefont {Kwon}, \citenamefont {Masson}, \citenamefont {Gutierrez-Jauregui}, \citenamefont {Asenjo-Garcia}, \citenamefont {Will},\ and\ \citenamefont {Yu}}]{huang_metasurface_2023}%
  \BibitemOpen
  \bibfield  {author} {\bibinfo {author} {\bibfnamefont {X.}~\bibnamefont {Huang}}, \bibinfo {author} {\bibfnamefont {W.}~\bibnamefont {Yuan}}, \bibinfo {author} {\bibfnamefont {A.}~\bibnamefont {Holman}}, \bibinfo {author} {\bibfnamefont {M.}~\bibnamefont {Kwon}}, \bibinfo {author} {\bibfnamefont {S.~J.}\ \bibnamefont {Masson}}, \bibinfo {author} {\bibfnamefont {R.}~\bibnamefont {Gutierrez-Jauregui}}, \bibinfo {author} {\bibfnamefont {A.}~\bibnamefont {Asenjo-Garcia}}, \bibinfo {author} {\bibfnamefont {S.}~\bibnamefont {Will}},\ and\ \bibinfo {author} {\bibfnamefont {N.}~\bibnamefont {Yu}},\ }\bibfield  {title} {\bibinfo {title} {Metasurface holographic optical traps for ultracold atoms},\ }\href {https://doi.org/10.1016/j.pquantelec.2023.100470} {\bibfield  {journal} {\bibinfo  {journal} {Progress in Quantum Electronics}\ }\textbf {\bibinfo {volume} {89}},\ \bibinfo {pages} {100470} (\bibinfo {year} {2023})}\BibitemShut {NoStop}%
\bibitem [{\citenamefont {Bar-Lev}\ \emph {et~al.}(2019)\citenamefont {Bar-Lev}, \citenamefont {England}, \citenamefont {Wootton}, \citenamefont {Liu}, \citenamefont {Gover}, \citenamefont {Byer}, \citenamefont {Leedle}, \citenamefont {Black},\ and\ \citenamefont {Scheuer}}]{bar-lev_design_2019}%
  \BibitemOpen
  \bibfield  {author} {\bibinfo {author} {\bibfnamefont {D.}~\bibnamefont {Bar-Lev}}, \bibinfo {author} {\bibfnamefont {R.~J.}\ \bibnamefont {England}}, \bibinfo {author} {\bibfnamefont {K.~P.}\ \bibnamefont {Wootton}}, \bibinfo {author} {\bibfnamefont {W.}~\bibnamefont {Liu}}, \bibinfo {author} {\bibfnamefont {A.}~\bibnamefont {Gover}}, \bibinfo {author} {\bibfnamefont {R.}~\bibnamefont {Byer}}, \bibinfo {author} {\bibfnamefont {K.~J.}\ \bibnamefont {Leedle}}, \bibinfo {author} {\bibfnamefont {D.}~\bibnamefont {Black}},\ and\ \bibinfo {author} {\bibfnamefont {J.}~\bibnamefont {Scheuer}},\ }\bibfield  {title} {\bibinfo {title} {Design of a plasmonic metasurface laser accelerator with a tapered phase velocity for subrelativistic particles},\ }\href {https://doi.org/10.1103/PhysRevAccelBeams.22.021303} {\bibfield  {journal} {\bibinfo  {journal} {Physical Review Accelerators and Beams}\ }\textbf {\bibinfo {volume} {22}},\ \bibinfo {pages} {021303} (\bibinfo {year} {2019})},\ \bibinfo {note} {publisher: American
  Physical Society}\BibitemShut {NoStop}%
\bibitem [{\citenamefont {Iványi}\ \emph {et~al.}(2024)\citenamefont {Iványi}, \citenamefont {Nemes}, \citenamefont {Gróf}, \citenamefont {Fekete}, \citenamefont {Kubacková}, \citenamefont {Tomori}, \citenamefont {Bánó}, \citenamefont {Vizsnyiczai},\ and\ \citenamefont {Kelemen}}]{ivanyi_optically_2024}%
  \BibitemOpen
  \bibfield  {author} {\bibinfo {author} {\bibfnamefont {G.~T.}\ \bibnamefont {Iványi}}, \bibinfo {author} {\bibfnamefont {B.}~\bibnamefont {Nemes}}, \bibinfo {author} {\bibfnamefont {I.}~\bibnamefont {Gróf}}, \bibinfo {author} {\bibfnamefont {T.}~\bibnamefont {Fekete}}, \bibinfo {author} {\bibfnamefont {J.}~\bibnamefont {Kubacková}}, \bibinfo {author} {\bibfnamefont {Z.}~\bibnamefont {Tomori}}, \bibinfo {author} {\bibfnamefont {G.}~\bibnamefont {Bánó}}, \bibinfo {author} {\bibfnamefont {G.}~\bibnamefont {Vizsnyiczai}},\ and\ \bibinfo {author} {\bibfnamefont {L.}~\bibnamefont {Kelemen}},\ }\bibfield  {title} {{\selectlanguage {english}\bibinfo {title} {Optically {Actuated} {Soft} {Microrobot} {Family} for {Single}-{Cell} {Manipulation}}},\ }\href {https://doi.org/10.1002/adma.202401115} {\bibfield  {journal} {\bibinfo  {journal} {Advanced Materials}\ }\textbf {\bibinfo {volume} {36}},\ \bibinfo {pages} {2401115} (\bibinfo {year} {2024})}\BibitemShut {NoStop}%
\bibitem [{\citenamefont {Bechinger}\ \emph {et~al.}(2016)\citenamefont {Bechinger}, \citenamefont {Di~Leonardo}, \citenamefont {Löwen}, \citenamefont {Reichhardt}, \citenamefont {Volpe},\ and\ \citenamefont {Volpe}}]{bechinger_active_2016}%
  \BibitemOpen
  \bibfield  {author} {\bibinfo {author} {\bibfnamefont {C.}~\bibnamefont {Bechinger}}, \bibinfo {author} {\bibfnamefont {R.}~\bibnamefont {Di~Leonardo}}, \bibinfo {author} {\bibfnamefont {H.}~\bibnamefont {Löwen}}, \bibinfo {author} {\bibfnamefont {C.}~\bibnamefont {Reichhardt}}, \bibinfo {author} {\bibfnamefont {G.}~\bibnamefont {Volpe}},\ and\ \bibinfo {author} {\bibfnamefont {G.}~\bibnamefont {Volpe}},\ }\bibfield  {title} {\bibinfo {title} {Active particles in complex and crowded environments},\ }\href {https://doi.org/10.1103/RevModPhys.88.045006} {\bibfield  {journal} {\bibinfo  {journal} {Reviews of Modern Physics}\ }\textbf {\bibinfo {volume} {88}},\ \bibinfo {pages} {045006} (\bibinfo {year} {2016})},\ \bibinfo {note} {publisher: American Physical Society}\BibitemShut {NoStop}%
\bibitem [{\citenamefont {Zemánek}\ \emph {et~al.}(2019)\citenamefont {Zemánek}, \citenamefont {Volpe}, \citenamefont {Jonáš},\ and\ \citenamefont {Brzobohatý}}]{zemanek_perspective_2019}%
  \BibitemOpen
  \bibfield  {author} {\bibinfo {author} {\bibfnamefont {P.}~\bibnamefont {Zemánek}}, \bibinfo {author} {\bibfnamefont {G.}~\bibnamefont {Volpe}}, \bibinfo {author} {\bibfnamefont {A.}~\bibnamefont {Jonáš}},\ and\ \bibinfo {author} {\bibfnamefont {O.}~\bibnamefont {Brzobohatý}},\ }\bibfield  {title} {{\selectlanguage {english}\bibinfo {title} {Perspective on light-induced transport of particles: from optical forces to phoretic motion}},\ }\href {https://doi.org/10.1364/AOP.11.000577} {\bibfield  {journal} {\bibinfo  {journal} {Advances in Optics and Photonics}\ }\textbf {\bibinfo {volume} {11}},\ \bibinfo {pages} {577} (\bibinfo {year} {2019})},\ \bibinfo {note} {publisher: Optica Publishing Group}\BibitemShut {NoStop}%
\bibitem [{\citenamefont {MacDonald}\ \emph {et~al.}(2003)\citenamefont {MacDonald}, \citenamefont {Spalding},\ and\ \citenamefont {Dholakia}}]{macdonald_microfluidic_2003}%
  \BibitemOpen
  \bibfield  {author} {\bibinfo {author} {\bibfnamefont {M.~P.}\ \bibnamefont {MacDonald}}, \bibinfo {author} {\bibfnamefont {G.~C.}\ \bibnamefont {Spalding}},\ and\ \bibinfo {author} {\bibfnamefont {K.}~\bibnamefont {Dholakia}},\ }\bibfield  {title} {{\selectlanguage {english}\bibinfo {title} {Microfluidic sorting in an optical lattice}},\ }\href {https://doi.org/10.1038/nature02144} {\bibfield  {journal} {\bibinfo  {journal} {Nature}\ }\textbf {\bibinfo {volume} {426}},\ \bibinfo {pages} {421} (\bibinfo {year} {2003})},\ \bibinfo {note} {publisher: Nature Publishing Group}\BibitemShut {NoStop}%
\bibitem [{\citenamefont {Shilkin}\ \emph {et~al.}(2017)\citenamefont {Shilkin}, \citenamefont {Lyubin}, \citenamefont {Shcherbakov}, \citenamefont {Lapine},\ and\ \citenamefont {Fedyanin}}]{shilkin_directional_2017}%
  \BibitemOpen
  \bibfield  {author} {\bibinfo {author} {\bibfnamefont {D.~A.}\ \bibnamefont {Shilkin}}, \bibinfo {author} {\bibfnamefont {E.~V.}\ \bibnamefont {Lyubin}}, \bibinfo {author} {\bibfnamefont {M.~R.}\ \bibnamefont {Shcherbakov}}, \bibinfo {author} {\bibfnamefont {M.}~\bibnamefont {Lapine}},\ and\ \bibinfo {author} {\bibfnamefont {A.~A.}\ \bibnamefont {Fedyanin}},\ }\bibfield  {title} {\bibinfo {title} {Directional {Optical} {Sorting} of {Silicon} {Nanoparticles}},\ }\href {https://doi.org/10.1021/acsphotonics.7b00574} {\bibfield  {journal} {\bibinfo  {journal} {ACS Photonics}\ }\textbf {\bibinfo {volume} {4}},\ \bibinfo {pages} {2312} (\bibinfo {year} {2017})},\ \bibinfo {note} {publisher: American Chemical Society}\BibitemShut {NoStop}%
\bibitem [{\citenamefont {Merrill}\ \emph {et~al.}(2009)\citenamefont {Merrill}, \citenamefont {Sainis},\ and\ \citenamefont {Dufresne}}]{merrill_many-body_2009}%
  \BibitemOpen
  \bibfield  {author} {\bibinfo {author} {\bibfnamefont {J.~W.}\ \bibnamefont {Merrill}}, \bibinfo {author} {\bibfnamefont {S.~K.}\ \bibnamefont {Sainis}},\ and\ \bibinfo {author} {\bibfnamefont {E.~R.}\ \bibnamefont {Dufresne}},\ }\bibfield  {title} {\bibinfo {title} {Many-{Body} {Electrostatic} {Forces} between {Colloidal} {Particles} at {Vanishing} {Ionic} {Strength}},\ }\href {https://doi.org/10.1103/PhysRevLett.103.138301} {\bibfield  {journal} {\bibinfo  {journal} {Physical Review Letters}\ }\textbf {\bibinfo {volume} {103}},\ \bibinfo {pages} {138301} (\bibinfo {year} {2009})},\ \bibinfo {note} {publisher: American Physical Society}\BibitemShut {NoStop}%
\bibitem [{\citenamefont {Chang}\ \emph {et~al.}(2018)\citenamefont {Chang}, \citenamefont {Douglas}, \citenamefont {González-Tudela}, \citenamefont {Hung},\ and\ \citenamefont {Kimble}}]{chang_colloquium_2018}%
  \BibitemOpen
  \bibfield  {author} {\bibinfo {author} {\bibfnamefont {D.}~\bibnamefont {Chang}}, \bibinfo {author} {\bibfnamefont {J.}~\bibnamefont {Douglas}}, \bibinfo {author} {\bibfnamefont {A.}~\bibnamefont {González-Tudela}}, \bibinfo {author} {\bibfnamefont {C.-L.}\ \bibnamefont {Hung}},\ and\ \bibinfo {author} {\bibfnamefont {H.}~\bibnamefont {Kimble}},\ }\bibfield  {title} {\bibinfo {title} {Colloquium: {Quantum} matter built from nanoscopic lattices of atoms and photons},\ }\href {https://doi.org/10.1103/RevModPhys.90.031002} {\bibfield  {journal} {\bibinfo  {journal} {Reviews of Modern Physics}\ }\textbf {\bibinfo {volume} {90}},\ \bibinfo {pages} {031002} (\bibinfo {year} {2018})},\ \bibinfo {note} {publisher: American Physical Society}\BibitemShut {NoStop}%
\bibitem [{\citenamefont {Jin}(2002)}]{jin_finite_2002}%
  \BibitemOpen
  \bibfield  {author} {\bibinfo {author} {\bibfnamefont {J.-M.}\ \bibnamefont {Jin}},\ }\bibfield  {title} {\bibinfo {title} {The {Finite} {Element} {Method} in {Electromagnetics}}\ }(\bibinfo {year} {2002})\BibitemShut {NoStop}%
\bibitem [{\citenamefont {Wang}\ \emph {et~al.}(2011)\citenamefont {Wang}, \citenamefont {Lazarov},\ and\ \citenamefont {Sigmund}}]{wang_projection_2011}%
  \BibitemOpen
  \bibfield  {author} {\bibinfo {author} {\bibfnamefont {F.}~\bibnamefont {Wang}}, \bibinfo {author} {\bibfnamefont {B.~S.}\ \bibnamefont {Lazarov}},\ and\ \bibinfo {author} {\bibfnamefont {O.}~\bibnamefont {Sigmund}},\ }\bibfield  {title} {{\selectlanguage {english}\bibinfo {title} {On projection methods, convergence and robust formulations in topology optimization}},\ }\href {https://doi.org/10.1007/s00158-010-0602-y} {\bibfield  {journal} {\bibinfo  {journal} {Structural and Multidisciplinary Optimization}\ }\textbf {\bibinfo {volume} {43}},\ \bibinfo {pages} {767} (\bibinfo {year} {2011})}\BibitemShut {NoStop}%
\bibitem [{\citenamefont {Jensen}\ and\ \citenamefont {Sigmund}(2005)}]{jensen_topology_2005}%
  \BibitemOpen
  \bibfield  {author} {\bibinfo {author} {\bibfnamefont {J.~S.}\ \bibnamefont {Jensen}}\ and\ \bibinfo {author} {\bibfnamefont {O.}~\bibnamefont {Sigmund}},\ }\bibfield  {title} {{\selectlanguage {english}\bibinfo {title} {Topology optimization of photonic crystal structures: a high-bandwidth low-loss {T}-junction waveguide}},\ }\href {https://doi.org/10.1364/JOSAB.22.001191} {\bibfield  {journal} {\bibinfo  {journal} {JOSA B}\ }\textbf {\bibinfo {volume} {22}},\ \bibinfo {pages} {1191} (\bibinfo {year} {2005})},\ \bibinfo {note} {publisher: Optica Publishing Group}\BibitemShut {NoStop}%
\bibitem [{\citenamefont {Li}\ \emph {et~al.}(2016)\citenamefont {Li}, \citenamefont {Chen}, \citenamefont {Liu},\ and\ \citenamefont {Tong}}]{li_structural_2016}%
  \BibitemOpen
  \bibfield  {author} {\bibinfo {author} {\bibfnamefont {Q.}~\bibnamefont {Li}}, \bibinfo {author} {\bibfnamefont {W.}~\bibnamefont {Chen}}, \bibinfo {author} {\bibfnamefont {S.}~\bibnamefont {Liu}},\ and\ \bibinfo {author} {\bibfnamefont {L.}~\bibnamefont {Tong}},\ }\bibfield  {title} {\bibinfo {title} {Structural topology optimization considering connectivity constraint},\ }\href {https://doi.org/10.1007/s00158-016-1459-5} {\bibfield  {journal} {\bibinfo  {journal} {Structural and Multidisciplinary Optimization}\ }\textbf {\bibinfo {volume} {54}},\ \bibinfo {pages} {971} (\bibinfo {year} {2016})}\BibitemShut {NoStop}%
\bibitem [{\citenamefont {Christiansen}(2023)}]{christiansen_inverse_2023}%
  \BibitemOpen
  \bibfield  {author} {\bibinfo {author} {\bibfnamefont {R.~E.}\ \bibnamefont {Christiansen}},\ }\bibfield  {title} {{\selectlanguage {english}\bibinfo {title} {Inverse design of optical mode converters by topology optimization: tutorial}},\ }\href {https://doi.org/10.1088/2040-8986/acdbdd} {\bibfield  {journal} {\bibinfo  {journal} {Journal of Optics}\ }\textbf {\bibinfo {volume} {25}},\ \bibinfo {pages} {083501} (\bibinfo {year} {2023})},\ \bibinfo {note} {publisher: IOP Publishing}\BibitemShut {NoStop}%
\bibitem [{\citenamefont {Svanberg}(1987)}]{svanberg_method_1987}%
  \BibitemOpen
  \bibfield  {author} {\bibinfo {author} {\bibfnamefont {K.}~\bibnamefont {Svanberg}},\ }\bibfield  {title} {{\selectlanguage {english}\bibinfo {title} {The method of moving asymptotes—a new method for structural optimization}},\ }\href {https://doi.org/10.1002/nme.1620240207} {\bibfield  {journal} {\bibinfo  {journal} {International Journal for Numerical Methods in Engineering}\ }\textbf {\bibinfo {volume} {24}},\ \bibinfo {pages} {359} (\bibinfo {year} {1987})}\BibitemShut {NoStop}%
\bibitem [{\citenamefont {Davoyan}\ \emph {et~al.}(2021)\citenamefont {Davoyan}, \citenamefont {Munday}, \citenamefont {Tabiryan}, \citenamefont {Swartzlander},\ and\ \citenamefont {Johnson}}]{Davoyan:21}%
  \BibitemOpen
  \bibfield  {author} {\bibinfo {author} {\bibfnamefont {A.~R.}\ \bibnamefont {Davoyan}}, \bibinfo {author} {\bibfnamefont {J.~N.}\ \bibnamefont {Munday}}, \bibinfo {author} {\bibfnamefont {N.}~\bibnamefont {Tabiryan}}, \bibinfo {author} {\bibfnamefont {G.~A.}\ \bibnamefont {Swartzlander}},\ and\ \bibinfo {author} {\bibfnamefont {L.}~\bibnamefont {Johnson}},\ }\bibfield  {title} {\bibinfo {title} {Photonic materials for interstellar solar sailing},\ }\href {https://doi.org/10.1364/OPTICA.417007} {\bibfield  {journal} {\bibinfo  {journal} {Optica}\ }\textbf {\bibinfo {volume} {8}},\ \bibinfo {pages} {722} (\bibinfo {year} {2021})}\BibitemShut {NoStop}%
\bibitem [{\citenamefont {Chen}\ \emph {et~al.}(2011)\citenamefont {Chen}, \citenamefont {Ng}, \citenamefont {Lin},\ and\ \citenamefont {Chan}}]{chenOpticalPullingForce2011}%
  \BibitemOpen
  \bibfield  {author} {\bibinfo {author} {\bibfnamefont {J.}~\bibnamefont {Chen}}, \bibinfo {author} {\bibfnamefont {J.}~\bibnamefont {Ng}}, \bibinfo {author} {\bibfnamefont {Z.}~\bibnamefont {Lin}},\ and\ \bibinfo {author} {\bibfnamefont {C.~T.}\ \bibnamefont {Chan}},\ }\bibfield  {title} {\bibinfo {title} {Optical pulling force},\ }\href {https://doi.org/10.1038/nphoton.2011.153} {\bibfield  {journal} {\bibinfo  {journal} {Nature Photonics}\ }\textbf {\bibinfo {volume} {5}},\ \bibinfo {pages} {531} (\bibinfo {year} {2011})}\BibitemShut {NoStop}%
\bibitem [{\citenamefont {Lu}\ \emph {et~al.}(2024)\citenamefont {Lu}, \citenamefont {Wen}, \citenamefont {Lu}, \citenamefont {Ding}, \citenamefont {Liu}, \citenamefont {Zheng},\ and\ \citenamefont {Chen}}]{Lu:24_pulling}%
  \BibitemOpen
  \bibfield  {author} {\bibinfo {author} {\bibfnamefont {L.}~\bibnamefont {Lu}}, \bibinfo {author} {\bibfnamefont {J.}~\bibnamefont {Wen}}, \bibinfo {author} {\bibfnamefont {M.}~\bibnamefont {Lu}}, \bibinfo {author} {\bibfnamefont {P.}~\bibnamefont {Ding}}, \bibinfo {author} {\bibfnamefont {J.}~\bibnamefont {Liu}}, \bibinfo {author} {\bibfnamefont {H.}~\bibnamefont {Zheng}},\ and\ \bibinfo {author} {\bibfnamefont {H.}~\bibnamefont {Chen}},\ }\bibfield  {title} {\bibinfo {title} {Interception force assisted optical pulling of a dipole nanoparticle in a single plane wave},\ }\href {https://doi.org/10.1364/OE.533355} {\bibfield  {journal} {\bibinfo  {journal} {Opt. Express}\ }\textbf {\bibinfo {volume} {32}},\ \bibinfo {pages} {31344} (\bibinfo {year} {2024})}\BibitemShut {NoStop}%
\bibitem [{\citenamefont {Jokisch}\ \emph {et~al.}(2024)\citenamefont {Jokisch}, \citenamefont {Christiansen},\ and\ \citenamefont {Sigmund}}]{jokisch_topology_2024}%
  \BibitemOpen
  \bibfield  {author} {\bibinfo {author} {\bibfnamefont {B.~M. d.~A.}\ \bibnamefont {Jokisch}}, \bibinfo {author} {\bibfnamefont {R.~E.}\ \bibnamefont {Christiansen}},\ and\ \bibinfo {author} {\bibfnamefont {O.}~\bibnamefont {Sigmund}},\ }\bibfield  {title} {{\selectlanguage {english}\bibinfo {title} {Topology optimization framework for designing efficient thermo-optical phase shifters}},\ }\href {https://doi.org/10.1364/JOSAB.499979} {\bibfield  {journal} {\bibinfo  {journal} {JOSA B}\ }\textbf {\bibinfo {volume} {41}},\ \bibinfo {pages} {A18} (\bibinfo {year} {2024})},\ \bibinfo {note} {publisher: Optica Publishing Group}\BibitemShut {NoStop}%
\end{thebibliography}%

\end{document}
%
% ****** End of file apssamp.tex ******